\documentclass{optica-article}

\journal{opticajournal} 

\articletype{Research Article}

\usepackage[section]{placeins}

\usepackage{braket}

\newcommand{\beginsupplement}{%
        \setcounter{table}{0}
        \renewcommand{\thetable}{S\arabic{table}}%
        \setcounter{figure}{0}
        \renewcommand{\thefigure}{S\arabic{figure}}%
     }

\begin{document}

\title{Multi-Plane Spatially Resolved Phase Structuring Using Optical Communication Modes}

\author{Vinicius S. de Angelis,\authormark{1,4*} Maximilian Jeindl,\authormark{2} Leonardo A. Ambrosio,\authormark{1}  David A. B. Miller,\authormark{3} Federico Capasso,\authormark{4} and Ahmed H. Dorrah \authormark{2,4 \dag}}
\address{\authormark{1}Department of Electrical and Computer Engineering, São Carlos School of Engineering, University of São Paulo, 400 Trabalhador São-Carlense Ave., 13566-590, São Carlos, São Paulo, Brazil \\
\authormark{2}Department of Applied Physics and Science Education, Eindhoven University of Technology, Eindhoven 5612 AP, The Netherlands\\
\authormark{3}Ginzton Laboratory, Stanford University, Stanford, CA 94305, USA \\
\authormark{4} Harvard John A. Paulson School of Engineering and Applied Sciences, Harvard University, Cambridge, MA 02138, USA \\
}

\email{\authormark{*} vinicius.angelis@usp.br  \hspace{0.5cm}\authormark{\dag} a.h.dorrah@tue.nl}

\begin{abstract*} 
We present a deterministic framework for three-dimensional beam shaping that enables versatile control of intensity and phase, pixel-by-pixel, across multiple axial planes. Conventional multi-plane holographic techniques typically rely on iterative optimization and mitigate inter-plane crosstalk through phase randomization, introducing speckle noise and thereby limiting deterministic phase control. Here, target fields are synthesized as a linear superposition of free-space communication modes obtained from the singular value decomposition of a coupling operator connecting a source plane to multiple target planes. Because these modes form orthogonal and energy-efficient transmission channels between the source and receiving spaces, their superposition yields volumetric wavefields with enforced phase coherence and reduced inter-plane crosstalk, without iterative refinement. We experimentally demonstrate high-fidelity reconstruction of intensity and phase profiles across multiple planes using a single phase-only spatial light modulator, including arbitrary structured phase singularity patterns. The proposed approach establishes communication-mode optics as a practical and physically grounded framework for multi-plane beam shaping, particularly in applications where phase structure and coherence across depth are essential.
\end{abstract*}

\section{Introduction}

Three-dimensional (3D) control of optical wavefields underpins applications in holography, optical manipulation, volumetric imaging, and diffractive optical systems \cite{Goodman2005, Piestun_2002, Grier2003, LAM2021050011,Huang2017,Lin2018}. While considerable progress has been made in synthesizing prescribed intensity distributions across multiple axial planes, deterministic control of phase distributions remains an outstanding challenge. Most multi-plane holographic techniques rely on iterative wave-propagation algorithms \cite{Gerchberg1972,Fienup1980,Tricoles:87,Olivier2004}, such as Global Gerchberg–Saxton (GGS) methods \cite{Haist,Sinclair:04} or stochastic gradient descent (SGD) optimization schemes \cite{Zhang:17,Chen:21_SGD}, to compute a single phase mask that reproduces target patterns at multiple depths. In these approaches, inter-plane crosstalk arises from the unavoidable overlap of reconstructed fields at adjacent planes, leading to ghost images and loss of contrast. Crosstalk suppression is commonly achieved by intentionally de-correlating the target planes, most notably by imposing random phase distributions \cite{Makey2019}. While random-phase encoding effectively suppresses coherent leakage between planes, it fundamentally precludes deterministic phase control and inevitably introduces speckle noise \cite{goodman2007speckle,Aageda1996}. Even advanced algorithms that explicitly penalize crosstalk and impose phase regularization, such as compensatory GGS \cite{Zhou:19} and double-constraint SGD methods \cite{Wang:23_cross_taslk_free}, remain incapable of eliminating all speckles --- an inherent limitation rooted in the phenomenon of vortex stagnation \cite{VortexStagnation}. As a result, existing multi-plane holographic techniques typically sacrifice phase determinism in exchange for intensity fidelity. This limitation is fundamental; namely, intensity-only multi-plane holography produces layered images rather than a single physically consistent wavefront.

On the other hand, phase consistency across axial planes is desirable as it ensures that the reconstructed fields correspond to a single physically valid wavefront, thereby preserving correct wavefront curvature and enabling genuine 3D depth cues \cite{Chang:20}. Beyond imaging, multi-plane phase control is also essential for compensating depth-dependent aberrations \cite{Li:19,Guo2025}, thereby enabling volumetric adaptive optics rather than a single global correction. Furthermore, the ability to control the phase profile at each plane allows the engineering of topological wavefields across planes, including phase singularities \cite{Berry2000-et,DENNIS2009,Shen2019-ss} that give rise to stacked on-demand blue-detuned optical trap arrays (for atom trapping) and genuinely volumetric trapping architectures, rather than isolated single-point traps \cite{Xu:10,Yu_2020,PhysRevLett_Barredo} or single-plane traps \cite{PhysRevA_Huft,PhysRevA_Piotrowicz,Ling_2023}. Finally, tailoring phase gradients enables the design of 3D optical force landscapes that support controlled rotation and circulation of particles, extending beyond the static trapping afforded by intensity gradients alone \cite{PhysRevLett.100.013602,Shi:23}. Nevertheless, a systematic framework that allows the phase profile to be arbitrarily structured, point-by-point, along the optical path remains elusive.

In this work, we propose and demonstrate a versatile method for controlling intensity and phase across multiple planes. The target field distributions are synthesized from a linear superposition of the communication modes connecting a source plane to the set of target planes. These modes are obtained using the singular value decomposition (SVD) modal optics \cite{Miller:19}. We encode the radiative near-field (Fresnel region) from the required source distribution into a CGH phase mask and optically reconstruct the resulting wavefront using a phase-only SLM. While communication mode theory \cite{Miller:98} originally established the foundations of SVD modal analysis, prior studies considered only theoretical scenarios in simplified geometries using prolate spheroidal functions \cite{Miller:00,Thaning_2003,Burvall:04}. A practical and experimentally validated framework for volumetric wavefront shaping with continuous depth of field has only recently been demonstrated \cite{deAngelis:25}. Here, we expand on this direction by focusing on multi-plane beam shaping and demonstrating on-demand, spatially-resolved, phase control.

Because the communication modes represent the optimal orthogonal channels between the source and receiving spaces, the corresponding source distribution generates an energy-optimal wavefront. This is expected to produce reconstructed intensity profiles with minimal crosstalk, high contrast and fidelity due to the efficient energy transfer among the target planes. Although iterative hologram synthesis can suppress speckle noise by initializing the target planes with well-defined phase distributions, notably constant and quadratic phase initializations \cite{theory_speckles,Shimobaba:15,Nagahama:19,Chen:21}, the choice of phase initialization influences both the convergence behavior and the achievable reconstruction accuracy, potentially converging to sub-optimal solutions, as the synthesis constitutes a non-convex optimization problem \cite{Sui2024-kw}. Our method, by contrast, is deterministic and requires no iterative refinement or inverse-design procedures. Moreover, since the communication modes are continuous complex-valued eigenfunctions, our method not only yields speckle-free intensity reconstructions but also enables the structuring of arbitrary phase distributions with high fidelity. To our knowledge, this work represents the first adaptation of SVD modal optics as a general and experimentally validated framework for multi-plane beam shaping of arbitrary complex-valued fields.

{\section{Concept}}

Each communication mode comprises a pair of eigenfunctions: one defined at the source space that leads to another at the receiving space. For scalar waves in free space, these eigenfunctions are connected to each other by a coupling operator $G_{SR}$ described by a scalar Green’s function. Modeling the spaces as being composed of sufficiently dense arrays of $N_S$ source and $N_R$ receiving points, $G_{SR}$ is expressed by a $N_S \times N_R$ matrix with its singular values $s_j$ establishing the connection between each pair of source $\ket{\Psi_{S,j}}$ and receiving $\ket{\Phi_{R,j}}$ eigenfunctions. In particular, these eigenfunctions satisfy the relationship $G_{SR} \ket{\Psi_{S,j}} = s_j \ket{\Phi_{R,j}}$, with $|s_j|^2$ quantifying their coupling strength. See Methods and Ref. \cite{Miller:19} for a detailed description of the communication mode theory.

\begin{figure}[htbp]
	\centering	\includegraphics[width=0.75\textwidth,height=1.0\textwidth]{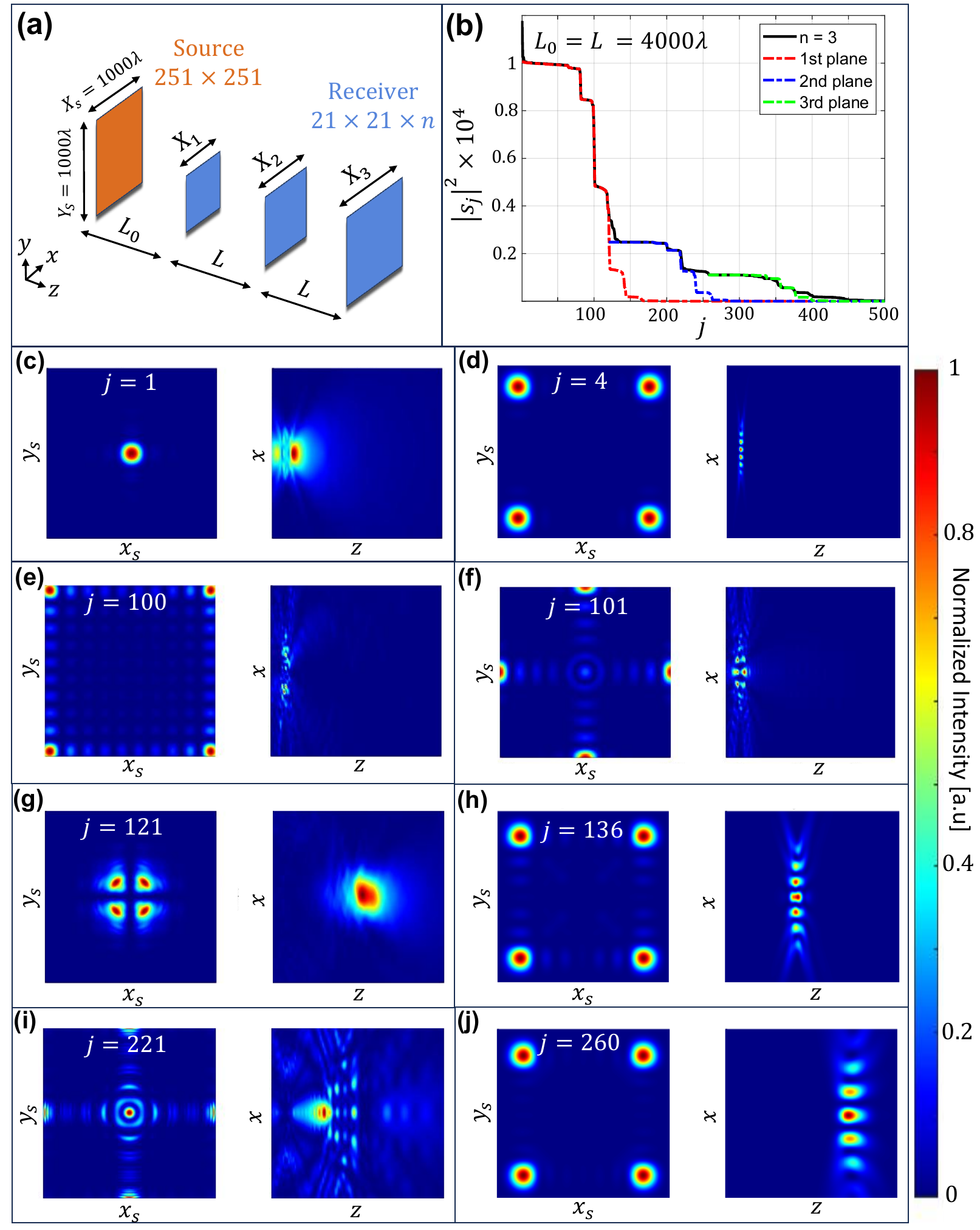}
	\caption{\small \textbf{Communication modes and their coupling strengths associated with a transverse source plane and a set of three transverse receiving planes equally spaced from each other}. \textbf{(a)} Dimensions of the source and receiving spaces, parametrized as listed in Table \ref{tab_param} for $n = 3$ receiving planes. \textbf{(b)} Coupling strengths in order of decreasing magnitude computed at $\lambda$ = 532 nm  (black solid line). For comparison, the coupling strengths associated with each receiving plane separately are also shown. Each $n$-th receiving plane supports a total of 100 strongly coupled modes, which are classified into two sub-categories: intrinsic and extrinsic modes. After the strong modes, each plane also supports partially coupled modes. Normalized squared amplitude of some modes. For the first plane: \textbf{(c)} $j = 1$ (extrinsic mode); \textbf{(d)} $j = 4$, \textbf{(e)} $j = 100$ (intrinsic modes); \textbf{(f)} $j = 101$ (partially coupled mode). For the second plane:  \textbf{(g)} $j = 121$ (extrinsic mode); \textbf{(h)} $j = 136$ (intrinsic mode), \textbf{(i)} $j = 221$ (partially coupled mode); For the third plane: \textbf{(j)} $j = 260$ (intrinsic mode). In each sub-figure the source eigenfunction is shown on the left and its resulting wave in the $xz$ plane on the right.}
\label{Fig_modes_n_2}
\end{figure}

For our multi-plane application, first we consider a source plane with transverse dimensions $X_s = Y_s$ and located at $z = 0$, as depicted in Fig. \ref{Fig_modes_n_2}(a). The receiving space consists of a set of $n$ transverse planes, uniformly spaced and without lateral shift; the distance from the source aperture to the $n$-th plane is $z_n = L_0 + (n-1) L$, where $L_0$ is the separation distance between the spaces. To remain in the paraxial regime we set $L_0 = 4 Y_s$. Each receiving plane is then sampled by an array of $p_{x,r} \times p_{y,r}$ receiving points, while the source plane is composed of $p_x \times p_y$ source points. We design the receiving planes such that all of them support the same number $N$ of strongly coupled modes. This leads to all the planes being subtended by a fixed solid angle from the source aperture with a linear increase of the spacing distances $d_{x,n} = d_{y,n}$ between the receiving points from plane to plane. The number $N$ of strongly coupled modes remains constant across the receiving planes and depends only on the receiving array. Finally, the spacing between the source points, $d_x = d_y$, is chosen such that wavefronts created by the $N_S$ source points converge to continuous functions (see Methods).

Table \ref{tab_param} lists the parameter values adopted for the configuration of Fig. \ref{Fig_modes_n_2}(a). With the chosen array size for the receiving points ($p_{x,r} = p_{y,r} = 21$), each receiving plane supports a total number of $N = 100$ strongly coupled modes. For operation at $\lambda$ = 532 nm, the coupling strengths $|s_j|^2$ in order of decreasing magnitude associated with these modes for $L = L_0$ are shown in Fig. \ref{Fig_modes_n_2}(b) (black solid line). For comparison, we also plot the coupling strengths obtained when each receiving plane is considered individually (red, blue, and green dashed lines for the first, second, and third planes, respectively). This comparison allows us to classify the strongly coupled modes associated with each $n$-th receiving plane into two sub-categories: (i) modes that are inherent to the $n$-th plane and thus are independent of the presence of the other receiving planes; and (ii) modes that arise due to the whole receiving space, influenced by the presence of the other receiving planes. We refer to modes of the first category as intrinsic and to those of the second as extrinsic. The coupling strengths of the intrinsic modes remain nearly constant over their respective ranges and decay from plane to plane according to the same inverse-square falloff as the intensity from an on-axis source point (see Supplementary Note 1).

\begin{table}[htbp]
\caption{Parameters adopted for the source and receiving spaces of Fig. \ref{Fig_modes_n_2}(a). Number of source and receiving points, spacing distances and longitudinal separation distance $L_0$ between the spaces.}
\centering
\label{tab_param}
\begin{tabular}{|cc|c|cc|l}
\cline{1-5}
\multicolumn{2}{|c|}{Source Plane}                   & \multirow{2}{*}{$L_0$} & \multicolumn{2}{c|}{Receiving space}                                                                                      &  \\ \cline{1-2} \cline{4-5}
\multicolumn{1}{|c|}{$p_x \times p_y$} & $d_x$,$d_y$ &                      & \multicolumn{1}{c|}{\begin{tabular}[c]{@{}c@{}}array of \\ points\end{tabular}} & spacing distances                       &  \\ \cline{1-5}
\multicolumn{1}{|c|}{$251 \times 251$} & $4\lambda$   & 4000$\lambda$         & \multicolumn{1}{c|}{$p_{x,r} \times p_{y,r} = 21 \times 21$}                  & $d_{x,n} = d_{y,n} = 2\lambda[1+(n-1)(L/L_0)]$ &  \\ \cline{1-5}
\end{tabular}
\end{table}

Although the number of intrinsic and extrinsic modes varies from plane to plane, their sum is constant: $N_{\text{int},n} + N_{\text{ext},n} = N$. In particular, the second (middle) plane exhibits the largest number of extrinsic strong modes. Figures \ref{Fig_modes_n_2}(c–j) show the normalized intensity distributions of selected modes at the source plane (left) and on the horizontal $xz$ plane (right). Supplementary Video 1 displays these distributions for the first 400 modes, including the intensity and phase at the three receiving planes. For the first plane, intrinsic behavior begins at $j = 4$ and extends to $j = 100$, implying $N_{\text{ext},1} = 3$ (i.e., $1 \leq j \leq 3$ are extrinsic modes). Modes $101 \leq j \leq 120$ form a partially coupled set not predicted by Eq. \ref{eq_heur_number_array_size}, with source-plane energy concentrated near the aperture edge. The strong modes of the second plane occupy the range $ 121 \leq j \leq 220$, with $N_{\text{ext},2} = 15$ extrinsic modes ($121 \leq j \leq 135$). Partially strong modes follow from $j = 221$. For the third plane, $N_{\text{ext},3} = 3$, with intrinsic behavior beginning at $j = 260$. Finally, after the partially strong modes ($336 \leq j \leq 358$), the coupling strengths fall off rapidly due to the universal tunneling-escape behavior of waves \cite{Miller2025-mm}, marking the onset of weakly coupled modes, with high energy and edge-localized source profiles.

\begin{figure}[htbp]
	\centering	\includegraphics[width=1.0\textwidth,height=1.0\textwidth]{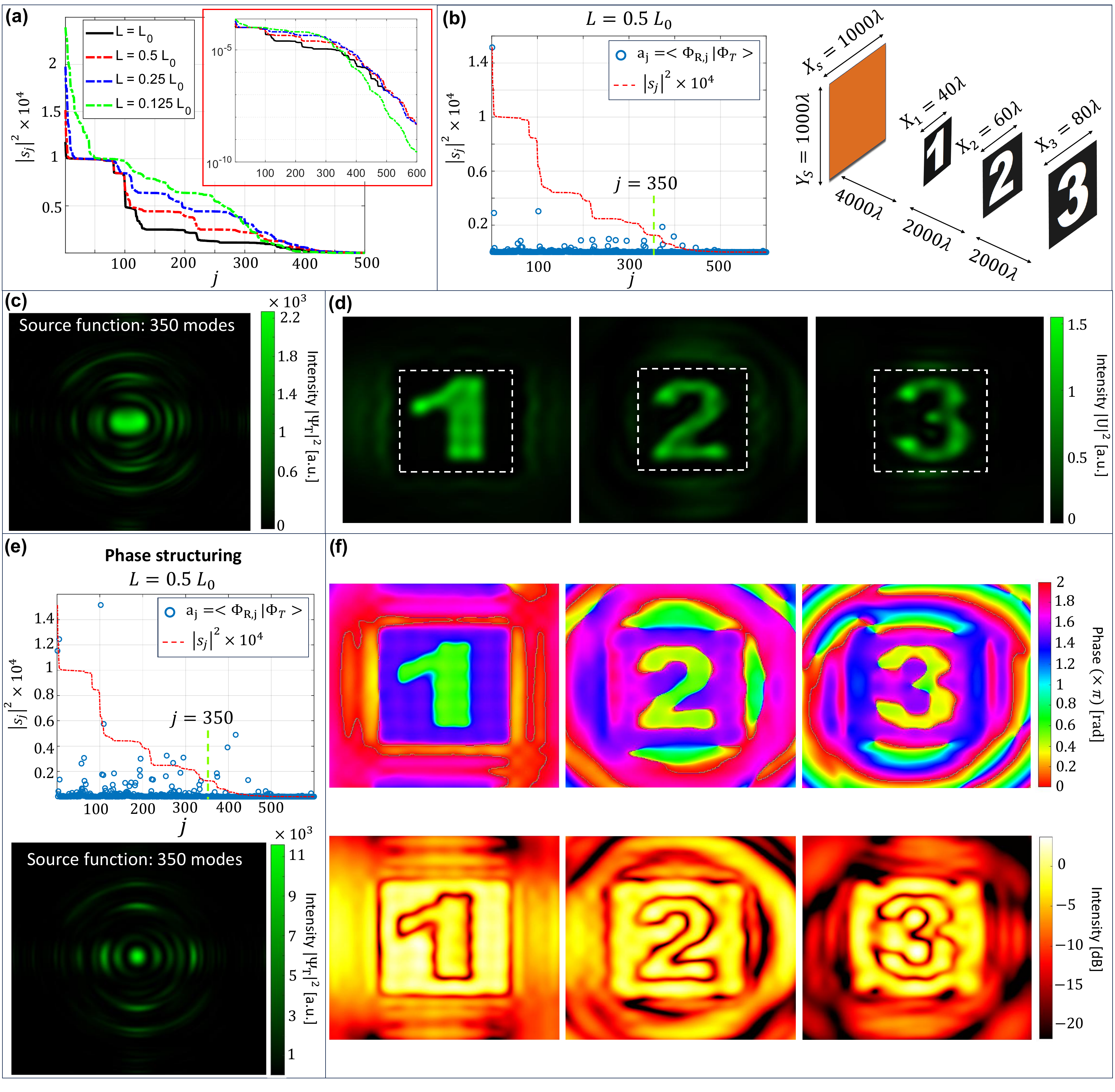}
	\caption{\small \textbf{Simulated structuring of arbitrary light wave profiles}. \textbf{(a)} Coupling strengths for different separation distances $L$ between the receiving planes in the configuration of Fig. \ref{Fig_modes_n_2}(a). The inset depicts the coupling strengths on a logarithmic scale. \textbf{(b)} For $L = 0.5L_0$, projection of a target intensity profile consisting of the binary digits '1', '2' and '3' onto the receiving basis set (blue circles). \textbf{(c)}  Calculated source function computed from the first 350 well-coupled modes (strong and partially coupled modes). \textbf{(d)} Calculated resulting wave from the source function in all three receiving planes. The dashed white squares highlight the size of the target planes. \textbf{(e)} Structuring phase profiles: the target consists of the same three binary digits with a phase of $\pi/2$ radians inside each digit’s contour and $3\pi/2$ radians outside. Amplitude is set to zero along each singularity contour (see Fig. S5). Projection onto the receiving basis set (blue circles) and calculated source function computed from the first 350 well-coupled modes (bottom). \textbf{(f)} Calculated phase and intensity (in dB) of the resulting wave in all three receiving planes.}
\label{Fig_svd_structuring_sim}
\end{figure}

As the separation distance between the receiving planes decreases, the number of extrinsic strong modes increases. This is illustrated in Fig. \ref{Fig_svd_structuring_sim}(a), which shows the coupling strengths for different values of the ratio $L/L_0$ in the configuration of Fig. \ref{Fig_modes_n_2}(a). Consequently, modes that originate from the whole receiving space become increasingly important for synthesizing light waves when the separation distances are reduced. However, upon further reduction of these distances, the total number of strongly coupled modes begins to decrease. This occurs for $L=0.125 L_0$, in which the number of strongly coupled modes in the second plane falls significantly below our design number of $100$. As a result, the coupling strengths decay much more rapidly than for larger values of $L/L_0$ (see the inset of Fig. \ref{Fig_svd_structuring_sim}(a), which shows the coupling strengths on a logarithmic scale), thereby limiting the number of usable modes available for structuring arbitrary light-wave profiles.

We next synthesize arbitrary structured light profiles in the receiving planes. First, the target profile is represented as a vector of complex amplitudes $\Phi_T$ at the receiving points and we project that profile onto the receiving basis. Finally, the required source function is computed as a linear superposition of the source components that create each receiving contribution of the target profile, akin to a Fourier series (see Methods). As an example, we consider as a target profile the intensity of binary digits '1,'2', and '3', assigned to each of the three receiving planes of the configuration of Fig. \ref{Fig_modes_n_2} for $L = 0.5 L_0$ -- see Fig. \ref{Fig_svd_structuring_sim}(b). The inner product coefficients between the receiving eigenfunctions and this target profile are shown in Fig. \ref{Fig_svd_structuring_sim}(b), depicted by the blue circles. For reference, the coupling strengths $|s_j|^2$ of this distribution are also shown in red dashed line. Using Eqs. \ref{eq_req_source} and \ref{eq_result_wave}, we compute the required source function (Fig. \ref{Fig_svd_structuring_sim}(c) shows the squared amplitude) using the first 350 well-coupled modes (strong and partially coupled modes). The intensity profile of the source function is depicted in Fig. \ref{Fig_svd_structuring_sim}(c) and the intensity distributions resulting from this source at all three receiving planes are shown in Fig. \ref{Fig_svd_structuring_sim}(d).

Incorporation of weakly coupled modes beyond the first 350 is impractical, as the gains in reconstruction accuracy afforded by these modes require an exponential increase in the source amplitudes \cite{Miller:19}. Owing to the tunneling-like escape of waves, the field generated by such high-energy source functions remains largely confined just outside the target planes, which results in inefficient reconstruction of the desired profiles. This behavior is illustrated in Supplementary Fig. S3, where the source function for the example of Fig. \ref{Fig_svd_structuring_sim}(b) is obtained from the first 500 coupled modes in an attempt to fully reconstruct the flat-topped profiles of the digits. A similar sub-diffraction regime arises when synthesizing target profiles for tightly spaced receiving planes, such as the case of $L=0.125L_0$. In this configuration, the set of strongly and partially coupled modes is insufficient to achieve a similar reconstruction accuracy as for larger $L/L_0$ values. As a result, a substantial number of weakly coupled modes must be included, which again yields a high-energy wave with the majority of its intensity localized at the edge of the receiving planes (see Supplementary Fig. S4).

Next, we structure arbitrary phase profiles in the receiving planes of the configuration in Fig. \ref{Fig_svd_structuring_sim}(b). As an example, we design two-dimensional (2D) phase-singularity sheets. In contrast to inverse design approaches which rely on maximizing the phase gradient at the desired singularity locations \cite{Lim2021-pc,Lim2023}, our method simply imposes target profiles with a $\pi$-phase discontinuity and dark intensity at those points. Here we assume the target phase profiles consisting of the digits '1', '2', and '3', defined with a phase of $\pi/2$ radians inside each digit’s contour and $3\pi/2$ radians outside. The target amplitude is set to zero along each singularity contour and to a uniform unit value elsewhere (see Supplementary Fig. S5). Figure \ref{Fig_svd_structuring_sim}(e) shows the projection of this target structure onto the receiving basis and the corresponding source field obtained from the first 350 well-coupled modes. The reconstructed phase and intensity (in dB) at the receiving planes are shown in Fig. \ref{Fig_svd_structuring_sim}(f). These fields represent the physically realizable wave emerging from the finite source aperture, with singular points reaching intensities as low as -24 dB across all planes. Approaching mathematically exact sheet singularities (with vanishing field amplitude) requires incorporating weakly coupled modes. Supplementary Fig. S6 shows the reconstructed profiles when the first 500 modes are incorporated, for which the minimum intensity at the singular points decreases to $-41$ dB.

\section{Experimental setup}

\begin{figure}[htbp]
	\centering	\includegraphics[width=1.0\textwidth,height=0.75\textwidth]{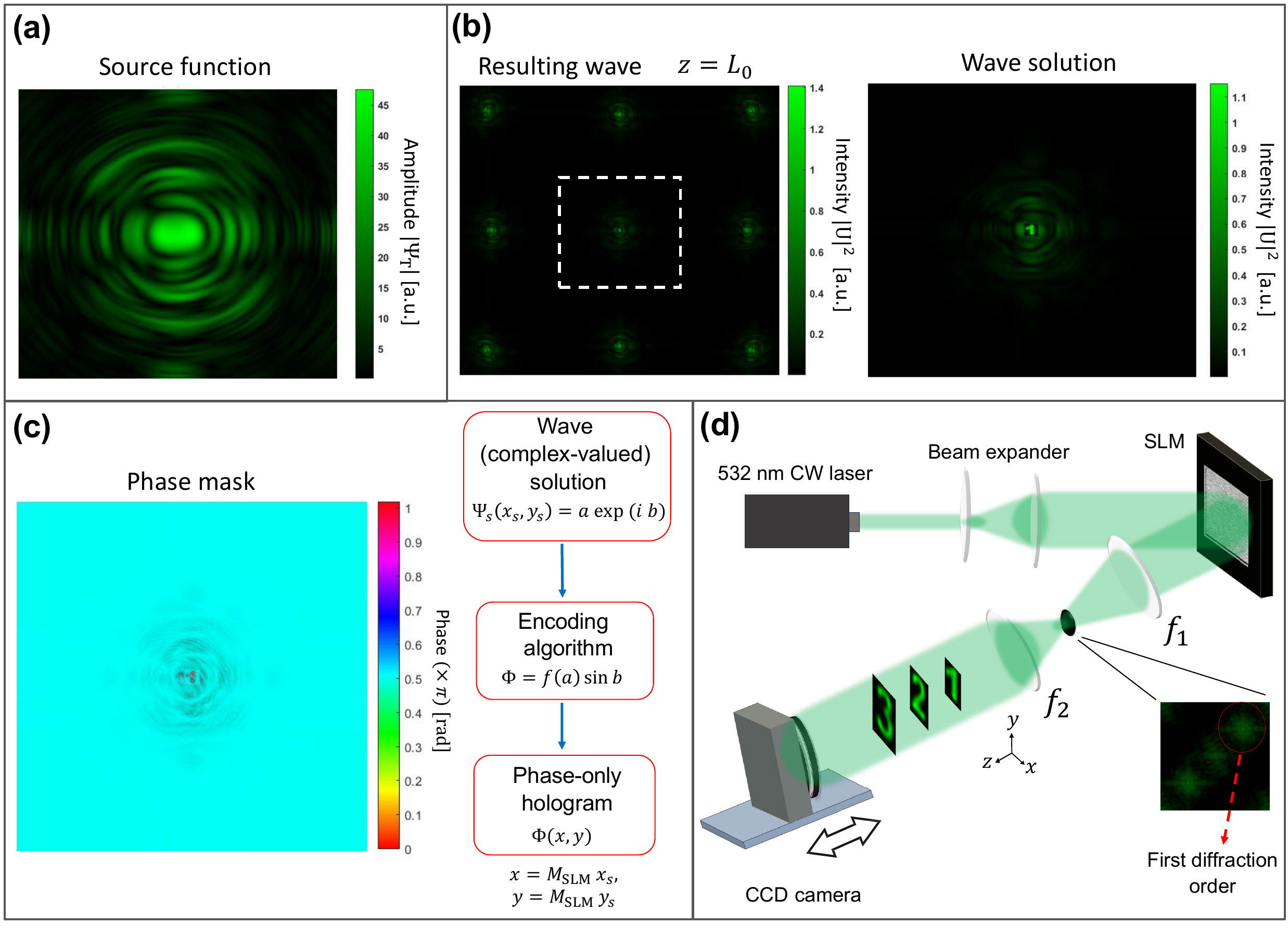}
	\caption{\small \textbf{Optical reconstruction of the light waves using a phase-only reflective SLM}. \textbf{(a)} Owing to its reactive near-field components and high amplitude source amplitudes originated from its diffraction orders, the source function cannot be directly encoded into a phase-only CGH. \textbf{(b)} The wave solution propagated from the source function to the plane $z = L_0$ was encoded instead as this Fresnel field is fully radiative and with no contributions from higher diffraction orders. The dashed white square highlights the aperture size. \textbf{(c)} The wave solution is interpolated to the utilized SLM display resolution ($801 \times 801$ pixels) and is converted into a phase mask using the deterministic hologram generation algorithm of Ref. \cite{Arrizon:07}. The algorithm encodes the wave solution on the first diffraction order of the phase mask in the spatial frequency domain. \textbf{(d)} Optical setup: a green laser is expanded and collimated before impinging on the SLM. After the SLM, a 4$f$ system is employed to recover the wave solution at the front focal plane of the second lens. An iris is placed at the Fourier plane to filter the first diffraction order while blocking the SLM unmodulated wave (zeroth diffraction order). The intensity distributions of the resulting wave are recorded using a CCD camera mounted on a translational stage. The phase profiles are reconstructed using the method of single-beam multiple-intensity reconstruction.}
\label{Fig_exp_setup}
\end{figure}

Several structured wave patterns were generated at $\lambda = $ 532 nm using a phase-only reflective spatial light modulator (SLM) with an 8 µm pixel pitch. The source function could not be directly encoded onto the SLM because it contains non-radiative (reactive) near-field components with transverse spatial frequencies exceeding the free-space propagation limit, and thus cannot be realized using a planar unitary phase modulator (see Supplementary Fig. S7). These components correspond to near-zone terms of the Green's function which arise due to the finite extent of the source and receiving planes. Additionally, the source function exhibits large amplitude values associated with diffraction orders, arising from the discrete sampling grid ($d_x=d_y>\lambda$), which are incompatible with the finite phase quantization of the SLM (10-bit resolution, 1024 phase levels). For these reasons, we instead encode the field propagated from the source plane to $z=L_0$. At this distance, the field contains only propagating components, with all evanescent contributions exponentially suppressed. Moreover, higher diffraction orders are spatially separated at $z=L_0$ and lie entirely outside the source aperture size, so that only the zeroth order is encoded; see, e.g., the dashed white square in Fig. \ref{Fig_exp_setup}(b).

To realize a complex field profile using a phase-only mask, several hologram generation algorithms have been proposed, with their performance typically evaluated in terms of reconstruction quality and diffraction efficiency \cite{Clark:16}. These methods rely on phase gratings to separate the shaped field -- usually encoded in the first diffraction order -- from the unshaped component. Here, we employ the algorithm of Ref. \cite{Arrizon:07} (CGH of Type 3), which provides high-fidelity recovery of the encoded complex field compared to alternative methods. A notable limitation of this approach is its relatively low diffraction efficiency, due to the limited modulation depth of the phase mask of only $1.17\pi$ radians, see Fig. \ref{Fig_exp_setup}(c), which reduces the intensity of the first-order diffracted field.

The experimental setup is depicted in Fig. \ref{Fig_exp_setup}(d). Before the SLM, the incident laser beam is collimated and expanded to illuminate the utilized SLM display area as a uniform quasi plane wave. A $4f$ imaging system is placed after the SLM to reconstruct the encoded complex field. At the Fourier plane an iris is placed to spatially filter the first diffraction order. Finally, the wave solution is projected at the front focal length of the second lens. In the encoding algorithm, the wave solution is interpolated to the utilized SLM display resolution ($801 \times 801$ pixels). Thus, to obtain a light field with the same transverse dimensions and longitudinal separations as those of the receiving planes in the configuration of Fig. \ref{Fig_svd_structuring_sim}(b), the aspect ratio of the $4f$ system must compensate for the SLM magnification $M_{\text{SLM}}$. This requires $M_{4f} = f_2/f_1 = 1/M_{\text{SLM}} = Y_s / (800 \times 8\mu) \approx 1/12.03$. However, to facilitate the measurements, we adopted $f_2/f_1 = 2/3$, which results in an $\times 8.02$ magnification for the transverse dimensions and a stretching of $\times 8.02^2 = \times 64.32$ in the longitudinal separation distances between the target planes. The propagated fields from the implemented CGHs were recorded using a CCD camera mounted on a translation stage as illustrated in Fig. \ref{Fig_exp_setup}(b). The phase profiles were retrieved using the single-beam multiple-intensity reconstruction (SBMIR) method \cite{Almoro:06}, in which the complex field at the plane of interest is reconstructed from a series of intensity measurements taken at multiple planes along the propagation direction without the use of a reference beam. For details, see Supplementary Note 2.

\section{Results}

\begin{figure}[htbp]
	\centering	\includegraphics[width=0.9\textwidth,height=1.2\textwidth]{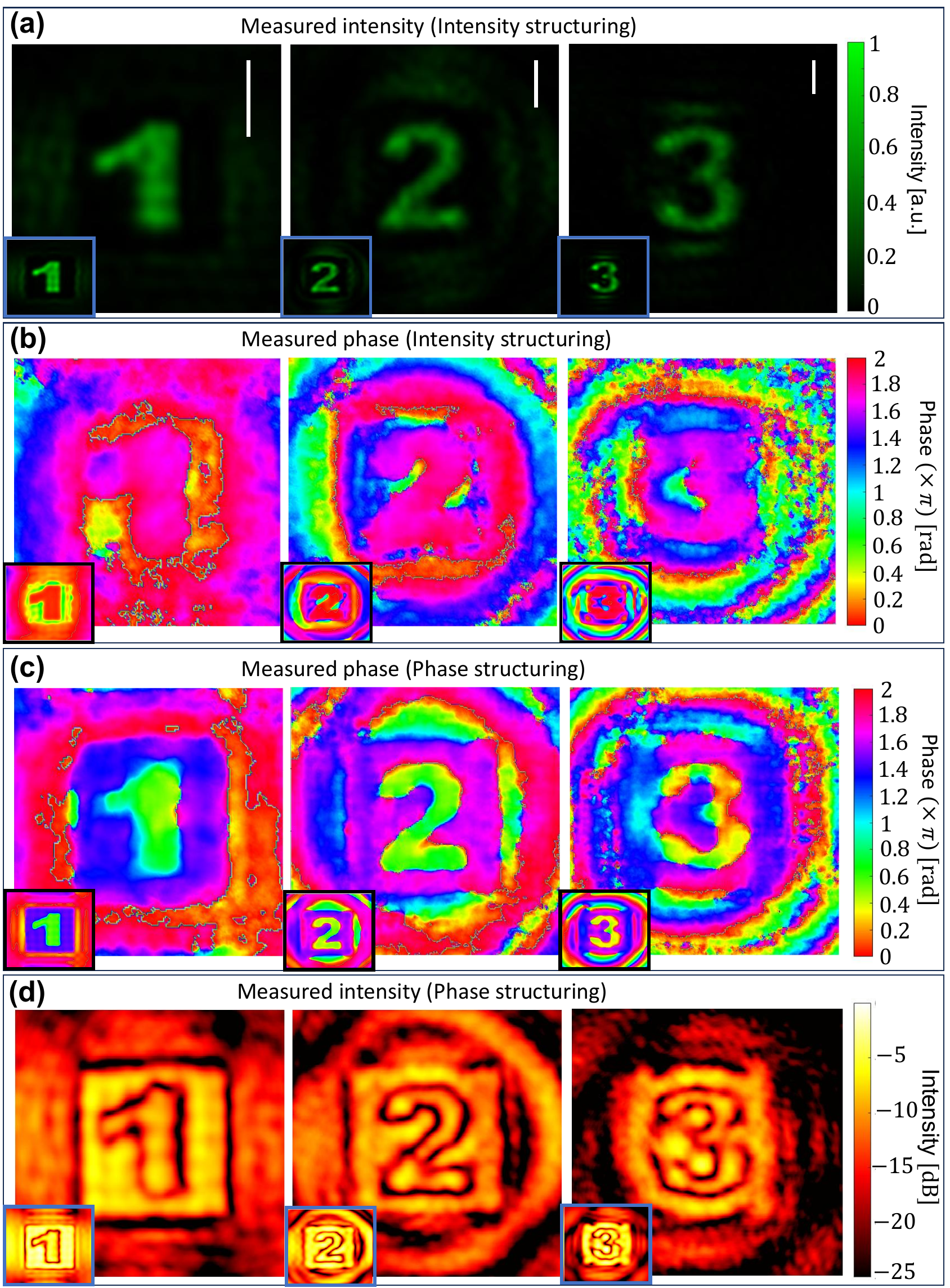}
	\caption{\small \textbf{Optical reconstruction of the light waves using a phase-only reflective SLM}. \textbf{(a)} Measured intensity and \textbf{(b)} retrieved phase distributions for the intensity-only target patterns of Fig. \ref{Fig_svd_structuring_sim}(d) at the three receiving planes. The reconstructed intensity profiles exhibit high fidelity, high contrast, and low speckle noise. \textbf{(c)} Measured retrieved phase and \textbf{(d)} intensity distributions for the target patterns of the example of Fig. \ref{Fig_svd_structuring_sim}(f). The digit's contours correspond to zero-amplitude lines where phase jumps of $\pi$ are imposed, forming two-dimensional phase-singularity sheets across all planes. Quantitative reconstruction metrics are reported in Tables 2 and 3. Scale bars (vertical white lines) represent 0.1 mm. The second and third planes are longitudinally spaced by 68.4 mm and 136.8 mm from the first plane. Inset figures show the simulated intensity and phase distributions.}
\label{Fig_results}
\end{figure}

We first experimentally reconstruct the multi-plane intensity-only case of Fig. \ref{Fig_svd_structuring_sim}(d). The measured intensity and retrieved phase distributions at the three target planes are shown in Figs. \ref{Fig_results}(a-b). To quantify the fidelity of the reconstructed profiles, we evaluated the mean squared error (MSE), signal-to-background ratio (SBR) and speckle contrast as defined in Methods. Table \ref{tab:metrics} summarizes these results. Across all planes, the measured intensity profiles closely match the target patterns with MSE below 0.1. The SBR exceeds 5.5 in every plane, indicating strong suppression of inter-plane crosstalk and the absence of visible ghosting artifacts. In addition, speckle contrast remains low, confirming that the deterministic superposition of communication modes yields locally smooth, speckle-free intensity profiles. Although no phase constraints are imposed in the intensity-only reconstructions, the recovered complex fields exhibit smooth and physically consistent phase distributions --- see Fig. \ref{Fig_results}(b) --- across all planes. Rather than exhibiting arbitrary or uncorrelated phase patterns, the reconstructed fields correspond to a single global solution of the scalar wave equation imposed by the superposition of the communication modes.

\begin{table}[ht]
\centering
\caption{Quantitative metrics for experimentally reconstructed intensity profiles in Fig. \ref{Fig_results}(a).}
\label{tab:metrics}
\begin{tabular}{c c c c c c}
\hline
Plane & MSE & SBR & Speckle Contrast\\
\hline
1  & 0.0452 & 11.68 & 0.139 \\
2  & 0.0729 & 9.31  & 0.231 \\
3  & 0.1006 & 5.70  & 0.357 \\
\hline
\end{tabular}
\end{table}

We next experimentally demonstrate the synthesis of 2D phase-singularity sheets across multiple planes. Figures \ref{Fig_results}(c-d) present the measured phase distributions and intensity (in dB) for the example of Fig. \ref{Fig_svd_structuring_sim}(f), showing that the prescribed zero-amplitude contours coincide with sharp $\pi$ phase transitions, forming continuous phase-singularity sheets that persist across all target planes. Quantitative phase reconstruction metrics are reported in Table \ref{tab:metrics_phase}. The MSE of the phase profiles remains low across all planes, indicating a high global phase fidelity. Furthermore, the standard deviation within the interior of each digit region ($\sigma_{\text{int}}$) and the outside region ($\sigma_{\text{out}}$) confirms good local phase uniformity away from the singularity contours. The measured minimum intensity at the singular points reaches values below -24 dB --- see Fig. \ref{Fig_results}(d), demonstrating the formation of physically realizable phase-singularity sheets with strong intensity suppression at the prescribed locations.

\begin{figure}[htbp]
	\centering	\includegraphics[width=1.0\textwidth,height=1.0\textwidth]{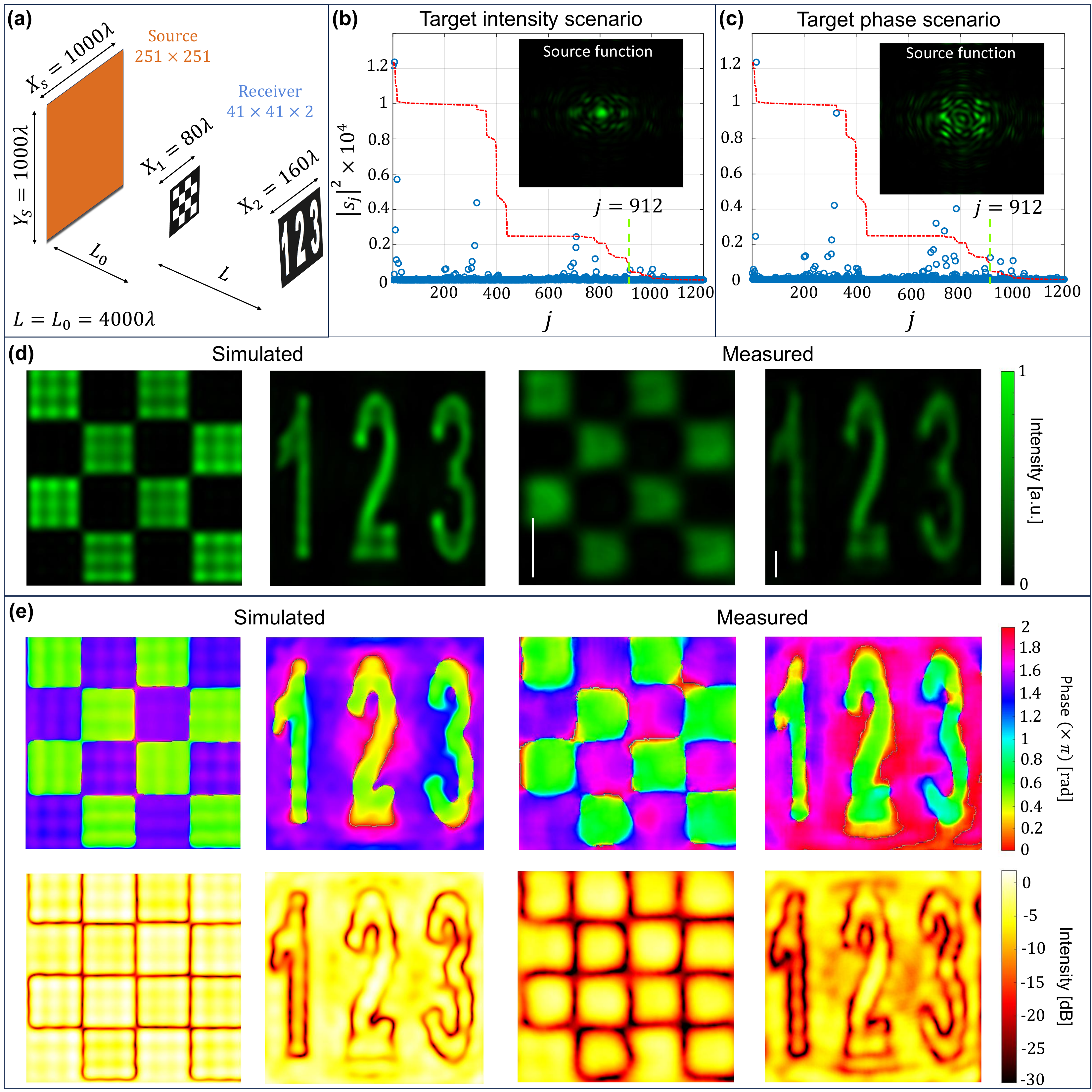}
	\caption{\small \textbf{Structuring high-resolution and multi-plane complex amplitude wavefront profiles}. \textbf{(a)} The number of supported strong modes in each receiving plane is increased by the array size. The same configuration of Fig. \ref{Fig_modes_n_2}(a) with $L = L_0$ and receiving planes composed of $41 \times 41$ receiving points, resulting in 400 strong modes in each plane. \textbf{(b)} Projection of the target patterns (checkerboard and sequence of digits '123') as intensity profiles and \textbf{(c)} as phase profiles. The inset figure shows the required source function computed from the first 912 well-coupled modes. Simulated resulting wave at the receiving planes and measured reconstructed profiles when the target patterns are \textbf{(d)} intensity-only profiles and \textbf{(e)} phase profiles (with the same target values as in Fig. \ref{Fig_svd_structuring_sim}). Scale bars (vertical white lines) represent 0.1 mm. The measured planes are longitudinally spaced by 136.8 mm.}
\label{Fig_higher_resolution}
\end{figure}

\begin{table}[ht]
\centering
\caption{Quantitative metrics for experimentally reconstructed phase profiles in Fig. \ref{Fig_results}(c).}
\label{tab:metrics_phase}
\begin{tabular}{c c c c c c}
\hline
Plane & MSE & $\sigma_{\text{int}} (\times \pi)$ & $\sigma_{\text{out}} (\times \pi)$\\
\hline
1  & 0.0712 & 0.1616 & 0.0708 \\
2  & 0.1172 & 0.1449  & 0.1153 \\
3  & 0.1924 & 0.4667  & 0.2953 \\
\hline
\end{tabular}
\end{table}

Now we apply the proposed method to the synthesis of higher-resolution patterns containing more stringent constraints in each target plane. As predicted by Eq. \ref{eq_heur_number_array_size}, the number of strong modes supported at each receiving plane increases with the array size of the receiving planes. Figure \ref{Fig_higher_resolution}(a) shows two receiving planes, each consisting of a $41 \times 41$ array of receiving points, supporting $N = 400$ strong modes per plane. Achieving this higher spatial resolution --- while preserving the source aperture size from Fig. \ref{Fig_modes_n_2}(a) and ensuring $N = 400$ strongly coupled modes -- imposes a constraint on the allowable separation between the target planes. The number of strong modes starts to decrease for separations smaller than $L = 0.25L_0$ (see Supplementary Fig. S8). At a separation distance of $L = L_0$, a checkerboard pattern is structured at the first receiving plane, while a pattern composed of the binary digits “123” is projected at the second plane. Figures \ref{Fig_higher_resolution}(b–c) show the projection of these target profiles onto the receiving basis, displayed as intensity and phase distributions using the same target values as in Fig. \ref{Fig_svd_structuring_sim}. The corresponding source functions (insets) are computed using the first 912 well-coupled modes, which accounts for all strong and partially coupled modes. Figures \ref{Fig_higher_resolution}(d–e) present the simulated wave fields at the receiving planes and the experimentally measured intensity and phase profiles using the SLM optical setup of Fig. \ref{Fig_exp_setup}, for both target profiles. Quantitative metrics for these measured profiles are presented in Table S2. Although a moderate increase in reconstruction error is observed relative to Fig. \ref{Fig_results}, the experimentally reconstructed intensity and phase profiles remain in good qualitative agreement with the designed targets across all planes, with the main structural features preserved. Additionally, the reconstructed phase distributions maintain coherent phase structure and sharp transitions at prescribed intensity minima, confirming that physically consistent complex-field solutions are enforced even in this more demanding scenario.

Additional experimental results illustrating the robustness of the proposed framework are presented in Supplementary Fig. S9. These measurements correspond to a configuration of three target planes in the non-paraxial regime as shown in Fig. S9(a). The same target phase profiles as in Fig. \ref{Fig_svd_structuring_sim} are assumed and the source function is synthesized from the first 250 well-coupled modes --- see Fig. S9(b).
Measured retrieved phase profiles are presented in Fig. S9(c) and these distributions remain in good agreement with the designed targets (MSE values shown in white). In the non-paraxial regime, the number of strongly coupled modes increases with propagation and reaches the heuristic value (Eq. \ref{eq_heuristic_number}) only near the paraxial regime; nevertheless, the field structure remains qualitatively intact and inter-plane crosstalk is strongly suppressed. These results confirm that the proposed approach provides a stable and predictable route to multi-plane field synthesis for distinct configurations of target planes.

\section{Conclusions}

We introduced and experimentally demonstrated a deterministic framework for 3D beam shaping that enables versatile control of intensity and phase across multiple axial planes. By expressing target fields as a linear superposition of free-space communication modes obtained from the singular value decomposition of a source–receiver coupling operator, the proposed approach constructs a single physically consistent wavefront rather than a set of independently reconstructed planes. Unlike iterative and inverse-design holographic techniques, which suppress inter-plane crosstalk by randomizing phase and thereby relinquish deterministic phase control, our method enforces phase coherence across planes by construction. This enables speckle-free reconstructions and direct synthesis of arbitrary phase profiles, including continuous topological structures such as 2D phase-singularity sheets, using a single static hologram. Additionally, the absence of iteration eliminates sensitivity to phase initializations and nonconvex convergence, providing robust solutions. Although the SVD calculation requires several hours for the configurations considered here (see Methods), this computation is performed once for a fixed source–receiver geometry. In contrast, iterative holographic algorithms require repeated forward–backward propagations for each new target field and must be re-executed for every design instance.

Beyond providing high fidelity reconstructions, the presented framework clarifies the physical limits of volumetric wave shaping with finite apertures. The communication modes define the complete set of volumetric fields that can be transmitted with high efficiency, revealing if certain field structures can be physically constructed or not based on the coupling strength of these modes. In this sense, SVD modal optics establishes a physically grounded basis for understanding and designing volumetric wave fields. In particular, inter-plane crosstalk is automatically eliminated through the linear superposition of the communication modes. Among these modes, the extrinsic ones — those imposed on each receiving plane by the presence of the other receiving planes — play a key role in suppressing coherent traces of adjacent-plane reconstructions. This effect is analyzed in Supplementary Fig. S10, where we show that excluding the extrinsic modes for the configuration in Fig. \ref{Fig_svd_structuring_sim}(b) leads to degraded reconstruction of the target intensity profiles. Furthermore, diffraction limits can be straightforwardly identified by computing the singular values for a given configuration of receiving planes and aperture size (i.e., a given coupling matrix $G_{SR}$) and comparing the number of strongly coupled modes with the total number of designed modes. This analysis reveals a minimum separation distance $L$ between planes: reducing $L$ below this threshold requires incorporating a substantial number of weakly coupled modes, which in turn demands excessively large source amplitude functions that are impractical to implement. Because the weak coupling strengths exhibit a universal rapid fall-off \cite{Miller2025-mm}, these limits are also universal; consequently, no other wave-shaping technique can yield physically realizable solutions in this sub-diffraction regime.

Deterministic control of phase enables a broad class of applications, including depth-dependent aberration correction, multi-plane structured illumination, and volumetric optical trapping based on phase-gradient forces. Beyond these examples, the proposed framework may extend to volumetric optical computing and diffractive neural networks, where programmable 3D operators require independent phase control per plane with reduced interlayer crosstalk. Multi-plane phase control further enables the engineering of constructive and destructive interference throughout the propagation volume, a capability central to nonlinear optical processes and quantum control schemes that rely on phase matching and coherent interaction across depths. More generally, the present results establish that SVD modal optics provides a physically consistent framework for multi-plane beam shaping in regimes where phase structure and coherence are important design considerations.

\begin{backmatter}
\bmsection{Acknowledgments}
V.S.A. acknowledges financial support from the Coordination of Superior Level Staff Improvement (CAPES), grant no. 88887.833874/2023-00, and from the National Council for Scientific and Technological Development (CNPq), grant no. 140270/2022-1. A.H.D. acknowledges financial support from the Dutch Research Council (NWO) under the Vidi program and from the Optica Foundation. L.A.A. acknowledges financial support from the National Council for Scientific and Technological Development (CNPq), grant no. 309201/2021-7, and from the São Paulo Research Foundation (FAPESP), grant no. 2025/28712-1. F.C. acknowledges financial support from the Office of Naval Research (ONR) under the MURI programme, grant no. N00014-20-1-2450, and from the Air Force Office of Scientific Research (AFOSR) under grant nos FA9550-21-1-0312 and FA9550-22-1-0243. D. M. also acknowledges support from AFOSR grant FA9550-21-1-0312.

\bmsection{Competing Interests}
The authors declare no competing interests.

\bmsection{Data Availability}
All key data supporting the findings of this study are included in the main article and its supplementary information. Additional data sets and raw measurements are available from the corresponding author upon reasonable request.

\bmsection{Supplemental document}
See Supplementary document for supporting content. 

\end{backmatter}

\bibliography{sample}

\section{Methods}

\subsection{Communication modes optics}



We assume the source and receiving spaces are Hilbert spaces, containing the possible source and receiving eigenfunctions $\ket{\Psi_{S}}$ and $\ket{\Phi_{R}}$. The coupling operator $G_{SR}$ between these spaces maps a position $\textbf{r}_S$ at the source space to a position $\textbf{r}_R$ at the receiving space for a given operating wavelength $\lambda$. For free-space scalar waves with time harmonic dependence $\exp(-\text{i}\omega_0 t)$, $G_{SR}$ is given by the usual free-space scalar Green's function for the Helmholtz equation \cite{hanson2001operator}:  
\begin{equation} \label{eq_GRS}
    G_{SR,\lambda}(\textbf{r}_R,\textbf{r}_S) = - \frac{1}{4 \pi} \frac{\text{exp}(\text{i} k |\textbf{r}_R-\textbf{r}_S|)}{|\textbf{r}_R-\textbf{r}_S|} \text{,}
\end{equation}
in which $k = 2\pi /\lambda$ is the wave number and $\omega_0 = k c$ is the operating angular frequency ($c$ is the light speed in free space). To describe $G_{SR}$ in a matrix form, we assume the source space as a collection of $N_S$ source points located at positions $\textbf{r}_{S,j}$ ($j=1,...,N_S$). Similarly, the receiving space is a collection of $N_R$ receiving points at positions $\textbf{r}_{R,i}$ ($i=1,...,N_R$). With this description, each eigenfunction, $\ket{\Psi_{S}}$ and $\ket{\Phi_{R}}$, is mathematically a column vector, whose elements are the complex-valued amplitudes at each different point in the appropriate space. This results in a $N_R \times N_S$ matrix for $G_{SR}$:
\begin{equation} \label{eq_GRS_matrix}
    g_{ij} = - \frac{1}{4 \pi} \frac{\text{exp}(\text{i} k |\textbf{r}_{R,i}-\textbf{r}_{S,j}|)}{|\textbf{r}_{R,i}-\textbf{r}_{S,j}|} \text{.}
\end{equation}

The concept of communication modes is established by the following two eigen-equations \cite{Miller:98,Miller:00,Miller:19}:  
\begin{subequations}\label{eq_eigenprob}
\begin{equation}
    G^{\dag}_{SR} G_{SR} \ket{\Psi_{S,j}}  = |s_j|^2 \ket{\Psi_{S,j}} \text{,}
\end{equation}
\begin{equation}
    G_{SR} G^{\dag}_{SR} \ket{\Phi_{R,j}} = |s_j|^2 \ket{\Phi_{R,j}} \text{,}
\end{equation}
\end{subequations}
in which $G^{\dag}_{SR} G_{SR}$ is a $N_S \times N_S$ matrix, i.e., an operator within the source space. Similarly, $G_{SR} G^{\dag}_{SR}$, with size $N_R \times N_R$, is an operator within the receiving space. Both are Hermitian and positive operators, implying that their eigenvalues are real positive numbers and that their eigenfunctions are orthogonal and form complete sets for the spaces. Additionally, the eigenvalues $|s_j|^2$, known as coupling strengths, correspond to the squared amplitude of the singular values $s_j$ of $G_{SR}$. In fact, from the two eigen-equations in Eq. \ref{eq_eigenprob} we obtain the following one-to-one relationship between the source and receiving eigenfunctions: 
\begin{equation}\label{eq_svd_mat_rel}
G_{SR} \ket{\Psi_{S,j}} = s_j \ket{\Phi_{R,j}} \text{.}
\end{equation}
The relation in Eq. \ref{eq_svd_mat_rel} defines the concept of a communication mode: a pair of eigenfunctions, one at the source space $\ket{\Psi_{S,j}}$ and one at the receiving space $\ket{\Phi_{R,j}}$, with the strength of this connection established by the coupling strength $|s_j|^2$. Each source eigenfunction $\ket{\Psi_{S,j}}$ creates a resulting wave in the receiving space that has the form of the receiving eigenfunction $\ket{\Phi_{R,j}}$ weighted by the complex amplitude $s_j$. Another property from the operators $G^{\dag}_{SR} G_{SR}$ and $G_{SR} G^{\dag}_{SR}$ is that their eigenfunctions and eigenvalues satisfy maximization properties, i.e., each successive receiving eigenfunction $\ket{\Phi_{R,j}}$ corresponds to the next largest possible magnitude of wave function its associated source eigenfunction $\ket{\Psi_{S,j}}$ can create at the receiving space \cite{Miller:19}.

\subsection{Design of the source and receiving planes}

 The spacing distances $d_{x,n}$ and $d_{y,n}$ in each receiving $n$-th plane must be designed based on the minimum spot length that can be structured from the source aperture along each transverse direction. In the paraxial regime, these quantities depend on the distance from the aperture $z_n = L_0 + (n-1) L$ as well as the aperture size ($X_s$ and $Y_s$) \cite{Piestun:96}:    
\begin{equation}
    \delta_{x,n} = \frac{\lambda z_n}{X_s} \hspace{0.5cm} \text{;} \hspace{0.5cm} \delta_{y,n} = \frac{\lambda z_n}{Y_s} \text{.}
\end{equation}

Setting the spacing distances $d_{x,n}$ and $d_{y,n}$ to satisfy the sampling rate according to the Nyquist theorem, i.e., $d_{x,n} = \delta_{x,n}/2$ and $d_{y,n} = \delta_{y,n}/2$, leads to:  
\begin{equation} \label{eq_spacing_dist_receiv_plane}
    d_{x,n} = \frac{\lambda}{2} \frac{z_n}{X_s}   \hspace{0.5cm} \text{;} \hspace{0.5cm}  d_{y,n} = \frac{\lambda}{2} \frac{z_n}{Y_s} \text{,}
\end{equation}
which implies a linear growth of the transverse dimensions of the receiving planes, $X_n = (p_{x,r,n}-1) d_{x,n}$ and $Y_n = (p_{y,r,n}-1) d_{y,n}$, with the distance from the source plane, in accordance with diffraction from a planar aperture \cite{Goodman2005}. The pyramidal volume defined by these dimensions represents the alias-free propagation region for an aperture sampled on a rectangular grid with spacing $d_x = d_y > \lambda$ \cite{Piestun:96}.




Now we evaluate the number of strong modes that each receiving plane can support using the paraxial heuristic number as defined in Refs \cite{Miller:00,Miller:19} between a source plane and a single receiving transverse plane. This number corresponds to the maximum number of intensity fringes that can be formed in the receiving plane: 
\begin{equation} \label{eq_heuristic_number}
    N_{H,n} = \frac{X_s X_n}{\lambda z_n} \frac{Y_s Y_n}{\lambda z_n} \text{.}
\end{equation}
 In the paraxial regime, the “Nyquist” heuristic counting, as given by Eq. \ref{eq_heuristic_number}, corresponds to the point at which the singular values from the communication modes analysis begin to fall off quasi-exponentially \cite{Miller:19}. Beyond the paraxial approximation, this simple heuristic number no longer applies, although the full communication modes analysis can still be used to determine the usable modes. Substituting the relations in Eq. \ref{eq_spacing_dist_receiv_plane} into Eq. \ref{eq_heuristic_number}, results in:
\begin{equation} \label{eq_heur_number_array_size}
    N_{H,n} = (1/4) (p_{x,r,n}-1) (p_{y,r,n}-1) \text{,}
\end{equation}
i.e., it corresponds to the number of points in the receiving array plane, so we are only using as many sampling points as we need for the number of strongly coupled modes. Setting the same number of receiving points for all the $n$ planes, i.e., $p_{x,r,n} = p_{y,r,n} = p_r$, all of them support the same amount of strong modes. From the relations in Eq. \ref{eq_spacing_dist_receiv_plane} and using that $X_s = Y_s = 4 L_0$, the spacing distances are written in terms of $(L_0/L)$ and $n$ as:
\begin{equation} \label{eq_spacing_distances_design}
    d_{x,n} = d_{y,n} = 2\lambda \Big[1 + (n-1) \frac{L}{L_0}\Big] \text{.}
\end{equation}


Therefore, we first specify the source dimensions, $X_s$ and $Y_s$, and determine the transverse spacing distances in the receiving planes according to Eq. \ref{eq_spacing_distances_design} for a given value for the ratio $L/L_0$. In all our examples in the main manuscript, we assumed $X_s = Y_s$ and $L_0 = 4Y_s$. Finally, the spacing distances in the source plane, $d_x = d_y$, are chosen such that the resulting wave created by the $N_s$ source points is essentially the same as if had a continuous source. In other words, we need to guarantee that the wave created by each source point is the same as a patch with the source point amplitude uniformly distributed over the area $d_x d_y$. To do this, we follow the procedure done in Ref. \cite{deAngelis:25}. First, for the largest distance between the source plane and each receiving plane, we compute the phase difference between the waves created by a point at the middle of the patch and one at the end of the patch (lying on the diagonal of the patch). Since destructive interference between these waves starts to occur if this phase difference is larger than $\lambda/2$, we then establish a criterion for the source spacings by setting this phase difference to be less than $\lambda/2$:
\begin{equation} \label{eq_cond_max_dx_dy}
   d_x = d_y < \frac{\lambda}{\sqrt{2}} \frac{\sqrt{(X_s+X_n)^2/2+z^2_n}}{(X_s+X_n)/2} \text{,} 
\end{equation}
for each receiving plane $n$. The criterion of Eq. \ref{eq_cond_max_dx_dy} is applied for $n=1$ since the first receiving plane imposes the smallest value for $d_x$. 



\subsection{Numerical computation of communication modes:} 

The communication modes were computed using a truncated singular value decomposition (svds, MATLAB) of the coupling matrix. For the configuration of Fig. \ref{Fig_svd_structuring_sim}(a), computation of the first 600 modes required 3 h 24 min, while the higher-resolution configuration of Fig. \ref{Fig_higher_resolution}(a) required 6 h 57 min for 1200 modes. The computation was performed without parallelization or GPU acceleration. Further reductions in runtime can be achieved using optimized or hardware-accelerated SVD implementations. This is a one-time offline computation for a fixed geometry of source aperture and receiving planes. 



\subsection{Synthesis of light waves based on communication modes}

The required source function to create a target complex-valued profile in the receiving space is \cite{Piestun_2002,Miller:19}: 
\begin{equation} \label{eq_req_source}
    \ket{\Psi_T} = \sum_{j} \frac{1}{s_j} \braket{\Phi_{R,j}|\Phi_T} \ket{\Psi_{S,j}}\text{,}
\end{equation}
in which $\braket{\Phi_{R,j}|\Phi_T}$ is the projection of the target profile onto the set of receiving eigenfunctions. 

The resulting wave created by this source function at a position $\textbf{r}$ from the source plane is given by summing up the spherical waves emitted by the $N_s$ source points weighted by $\ket{\Psi_T}$ \cite{Miller:19}:
\begin{equation} \label{eq_result_wave}
    U(\textbf{r}) = - \frac{1}{4 \pi} \sum_{q=1}^{N_S} \frac{\text{exp}(i k |\textbf{r}-\textbf{r}_{S,q}|)}{|\textbf{r}-\textbf{r}_{S,q}|} h_q \text{,}
\end{equation}
where $h_q$ is the $q$-th component of the source function $\ket{\Psi_T}$ on the source basis $\{\ket{\Psi_{S,j}}\}$. 


\subsection{Reconstruction quality metrics}


\textbf{Mean squared error (MSE):} 
\begin{equation}
    \text{MSE} = \frac{1}{N} \sum_{i=1}^{N} \Big[f_{\text{meas}}(i)-f_{\text{targ}}(i) \Big]^2 \text{,}
\end{equation}
where the summation is taken over the $N$ pixels of the CCD camera within the target plane. All intensity distributions were normalized to their respective maximum values within each plane prior to computing MSE. Similarly, for phase reconstructions, the phase $\phi$ was first normalized as $(\phi-0.5)/\pi$.  



\textbf{Signal-to-background ratio (SBR):} 
\begin{equation}
    \text{SBR} = \frac{<I_{\text{sig}}>}{<I_{\text{bg}}>} \text{,}
\end{equation}
where $<I_{\text{sig}}>$ denotes the mean intensity within regions corresponding to the target signal (bright  regions), and $<I_{\text{bg}}>$ is the mean intensity within nominally dark background regions. The signal and background regions were defined based on the target intensity distribution using a fixed intensity threshold of 10\% of the maximum intensity.

\textbf{Speckle contrast:}
Local intensity fluctuations within nominally uniform bright regions were characterized using speckle contrast:
\begin{equation}
    C = \frac{\sigma_I}{<I>}  \text{,}
\end{equation}
in which $\sigma_I$ and $<I>$ are the standard deviation and mean of the intensity evaluated over the signal regions. Edge regions were excluded by morphological erosion to avoid bias from intensity gradients.


\subsection{Experimental setup}




As light source an Oxxius L1C+-532S 532.3 nm diode pumped solid state laser was used in conjunction with a pinhole spatial filter setup (horizontal polarization ensured by $\lambda/2$-plate and Thorlabs LPVIS050-MP2 polarizer, 30$\mu$m pinhole, 20x objective, 750 mm collimating plano-convex lens) to generate a well collimated Gaussian beam of sufficient diameter. Collimation was confirmed using a Thorlabs WFS20 Shack-Hartman Wavefront sensor. The radius of curvature was over 100 m.

The CGHs were generated using a Santec SLM-200 at normal incidence combined with a $4f$ system of magnification 
$M_{4f} = 100/150 = 2/3$ for Fourier-order filtering, employing plano-convex lenses. A non-polarizing cube-beam-splitter was used to couple the SLM output into the $4f$ Fourier filter, where a diffraction order was chosen using an iris. Longitudinal distance between the $4f$ lenses was also optimized to a radius of curvature over 50 m with an incident flat wavefront. The second lens of the $4f$ Fourier filter was shifted laterally to accommodate the chosen Fourier order.

Intensities of the CGHs were measured using a monochromatic IDS U3-3880CP-M-GL Rev.2.2 CMOS camera (2.4 $\mu$m pixel size) mounted on a Thorlab LTS150C/M linear stage. The total magnification of the system, product of the $4f$ magnification and the SLM magnification due to rescaling compared to the SVD design geometry, $M_{\text{SLM}} M_{4f}$, was chosen to fit the 150 mm of stage travel, so the full 3D space of interest could be measured at once.

\newpage

\section*{Supplementary Document}
\beginsupplement

\subsection*{Supplementary Note 1: coupling strengths of intrinsic modes}

\begin{figure}[htbp]
	\centering	\includegraphics[width=1.0\textwidth,height=0.45\textwidth]{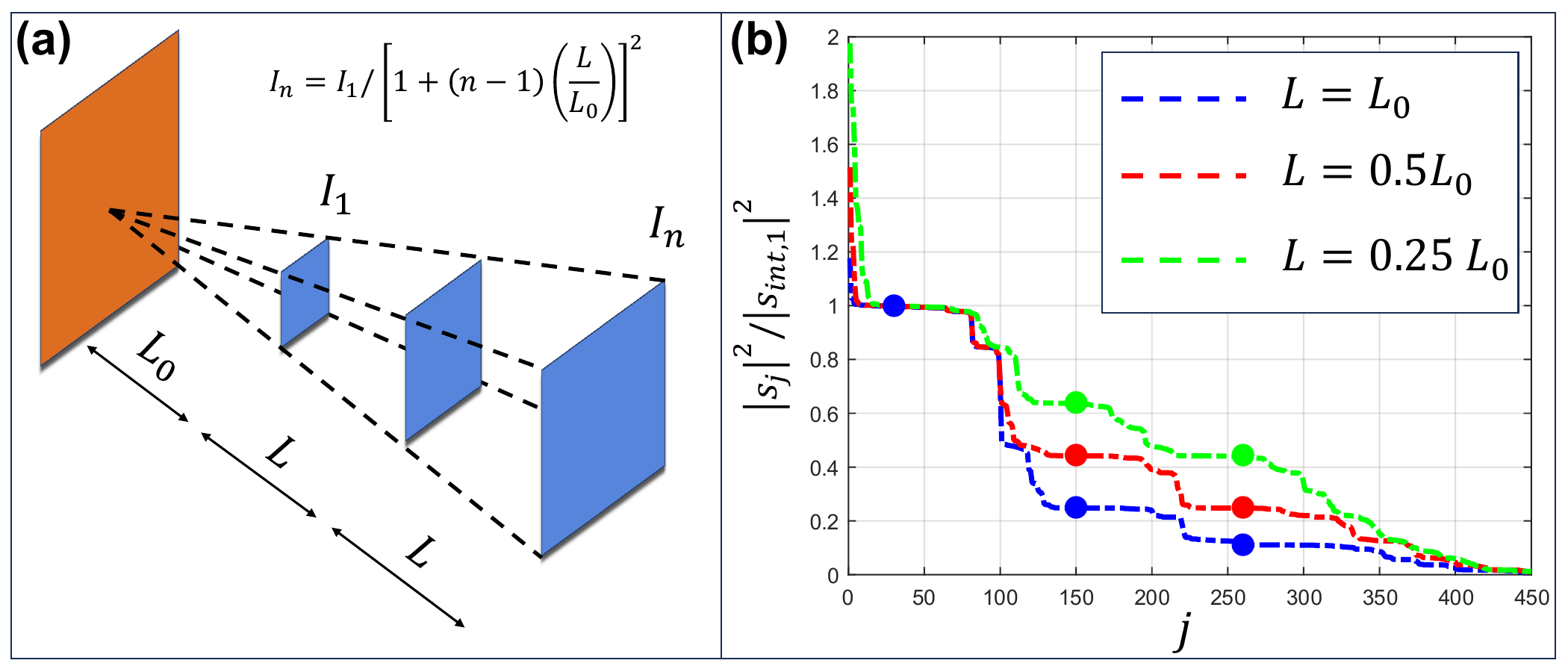}
	\caption{\small \textbf{Coupling strengths of intrinsic modes}. \textbf{(a)} Intensity decay from an on-axis source point centered at the source plane. \textbf{(b)} The coupling strengths of intrinsic modes also decays with the same inverse-square dependence. The dots represent the modal strengths computed from this dependence (see Eq. \ref{eq_int_coupling}).}
\label{Fig_suppl_intris_coup_str}
\end{figure}

Assume an on-axis source point centered at the source plane emitting a light wave with power $P$ toward the receiving planes as shown in Fig. \ref{Fig_suppl_intris_coup_str}(a). The light intensity at the $n$-th receiving plane, which has an area $A_n = X_n Y_n$, is: 
\begin{equation} \label{eq_int_decay}
    I_n = \frac{P}{A_n} = \frac{P}{(p_r-1)^2 4 \lambda^2[1+(n-1)(L/L_0)]^2} = \frac{I_1}{[1+(n-1)(L/L_0)]^2} \text{,}
\end{equation}
in which $I_1 = P/[4 \lambda^2(p_r-1)^2]$ is the light intensity at the first plane. The coupling strengths of the intrinsic modes decay from plane to plane with the same inverse-square dependence as the intensity, i.e.: 
\begin{equation} \label{eq_int_coupling}
    |s_{\text{int},n}|^2 = \frac{|s_{\text{int},1}|^2}{[1+(n-1)(L/L_0)]^2} \text{,}
\end{equation}
as shown in Fig. \ref{Fig_suppl_intris_coup_str}(b) for different values of $L/L_0$, in which the dots represent the values evaluated from Eq. \ref{eq_int_coupling}. This indicates that these modal strengths are simply proportional to the light intensities prescribed in Eq. \ref{eq_int_decay}.

\newpage

\subsection*{Supplementary Note 2: Phase retrieval algorithm}

A variation of the SBMIR algorithm by Almoro and Pedrini \cite{Almoro:06} was implemented in Matlab. A Fresnel propagation formalism was used to propagate the complex field-amplitudes between different planes. The SBMIR reconstructs the phase and amplitude of fields along propagation direction using intensity measurements acquired at $N$ planes. High and low resolution ranges in propagation direction were defined, where areas close to the designed planes have more measurement points. The recorded frame size was chosen to encompass the region containing features reasonably above the noise floor. Additionally, 12 frames were binned into a single measurement frame, and each datapoint consists of several such groups of measurement frames. An averaged dark frame of the camera at the same camera settings was subtracted from measurements before the phase retrieval. Pixel values that are negative after background subtraction are set to zero to avoid unphysical intensities.

\begin{figure}[hbtp]
\centering
\includegraphics[width=1.0\linewidth,height=0.6\textwidth]{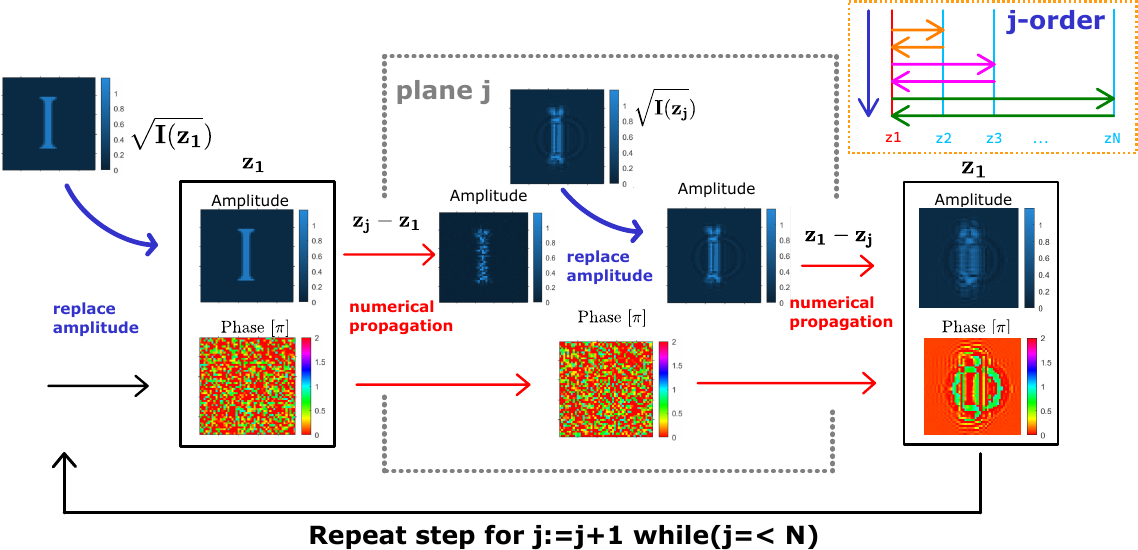}
\caption{Sub-step of SBMIR iteration step. Here $z_1$ is the first plane along the propagation direction and $z_j < z_{j+1}$ for all $j$. The inset on top right indicates the order in which the planes are included within a single iteration along the incremental propagation direction.}
\label{fig:SBMIRcombined}
\end{figure}

First, a starting plane with measured intensity $I_k$ is selected, and the corresponding complex field is initialized as ${A_k} = \sqrt{I_k}\cdot\exp\left(i\Phi_{k,\mathrm{initial}}\right)$, with $\Phi_{k,\mathrm{initial}}$ chosen as a pixel-wise random phase. Afterwards, the iterative algorithm starts with the assumed phase and amplitude $A_k$ numerically propagated a distance $z_m-z_k$ to a different measurement plane, as illustrated in Fig. \ref{fig:SBMIRcombined}. Here the field amplitude $A_m \rightarrow \sqrt{I_m}\cdot\exp\left(i\Phi_m \right)$ is replaced using the measured intensity, while the propagated phase $\Phi_m=\mathrm{angle}(A_{m,\mathrm{propagated}})$ is kept. Next, $A_m$ is propagated back $z_k-z_m$ to the position of the reference plane, where again the amplitude is replaced while the phase is kept. This yields the phase-amplitude $A_k = \sqrt{I_k}\cdot \exp\left(i\Phi_{k, \mathrm{propagated}}\right)$. For the next step, a new plane at $z_l$ is selected and the procedure is repeated until all planes of interest have been included, completing a single iteration. A fixed number of iterations is performed, after which convergence is assessed. The iteration counts for the different measurements, along with selected quality metrics, are reported in Table \ref{tab:valsPhaseRetrieval}.

\begin{table}[]
\caption{Phase retrieval metrics for each result measurements. RMS is the root mean square error comparing intensity measurement and retrieved squared amplitude. $I_{\text{max}}$ is the maximum pixel value over all planes. $N_{\text{iter}}$ is the number of SBMIR iterations performed. HR and LR refer to the adopted high and low resolution ranges in propagation direction. The former were set to areas close to the designed planes. Start and end indicate the index of the first and the last plane used in the phase retrieval. The start index always is the highest index and corresponds to the first measured plane in propagation direction. $d_{\text{meas}}$ is the total propagation distance covered by the measurement and $d_{\text{SBMIR}}$ is the percentage of that which lies between start and end index used for the phase retrieval. $\mathrm{N_F}$ is the number of frames per measurement point.
\label{tab:valsPhaseRetrieval}}
\begin{tabular}{l|r|r|r|l|r|r|r|r}
Measurement & $\frac{\mathrm{RMS}}{I_{\text{max}}}$ / a.u. & $N_{\text{iter}}$ & \begin{tabular}[c]{@{}l@{}}HR (LR)\\ (mm) \end{tabular} & \begin{tabular}[c]{@{}l@{}}$d_{\text{meas}}$\\ (mm) \end{tabular}
 & start & end & \begin{tabular}[c]{@{}l@{}}$d_{\text{SBMIR}}$\\ (\%) \end{tabular} & $\mathrm{N_{F}}$ \\ \hline
Figs. \ref{Fig_results}(a-b)  & 0.0127                              & 400        & 0.2 (1.0)    & 150             & 287   & 120 & 53               & 36               \\
Figs. \ref{Fig_results}(c-d)  & 0.0104                              & 800        & 0.2 (1.0)    & 150             & 287   & 120 & 71               & 48               \\
Fig. \ref{Fig_higher_resolution}(d)  & 0.0101                              & 800        & 0.2 (1.0)    & 150             & 247   & 100 & 53               & 36               \\
Fig. \ref{Fig_higher_resolution}(e)  & 0.0123                              & 800        & 0.2 (1.0)    & 150             & 247   & 100 & 71               & 36               \\
Fig. \ref{Fig_non_paraxial_meas}    & 0.00786                              & 800        & 0.25 (0.25)  & 102             & 409   & 190 & 54               & 12              
\end{tabular}
\end{table}

The first designed plane, which lies at the focal distance of the second lens of the $4f$ system, was chosen as the reference plane. Each sub-cycle starts and ends at that plane as shown in Fig. \ref{fig:SBMIRcombined}. The next planes were chosen incrementally along the propagation direction. Choosing the first plane as a reference was done for predictability, even though it implies a higher weight towards the first plane for the reconstructed fields. For computational efficiency, the iterations, which rely heavily on Fast-Fourier-Transforms, were done on a GPU. Due to memory limitations, only a part of the measurement could be included in the iterations. Using the first approximately 100 planes, along the propagation direction, for the phase retrieval yielded more stable convergence than including only every second plane of the entire propagation range or taking the last 100 planes. The result of the phase retrieval was manually evaluated after a fixed number of iterations and is summarized in Table \ref{tab:valsPhaseRetrieval}. Supplementary Video 2 presents the measured intensity and retrieved phase of all measured planes for the phase structuring case shown in Figs. \ref{Fig_results}(c-d).

\begin{figure}[htbp]
	\centering	\includegraphics[width=1.0\textwidth,height=0.85\textwidth]{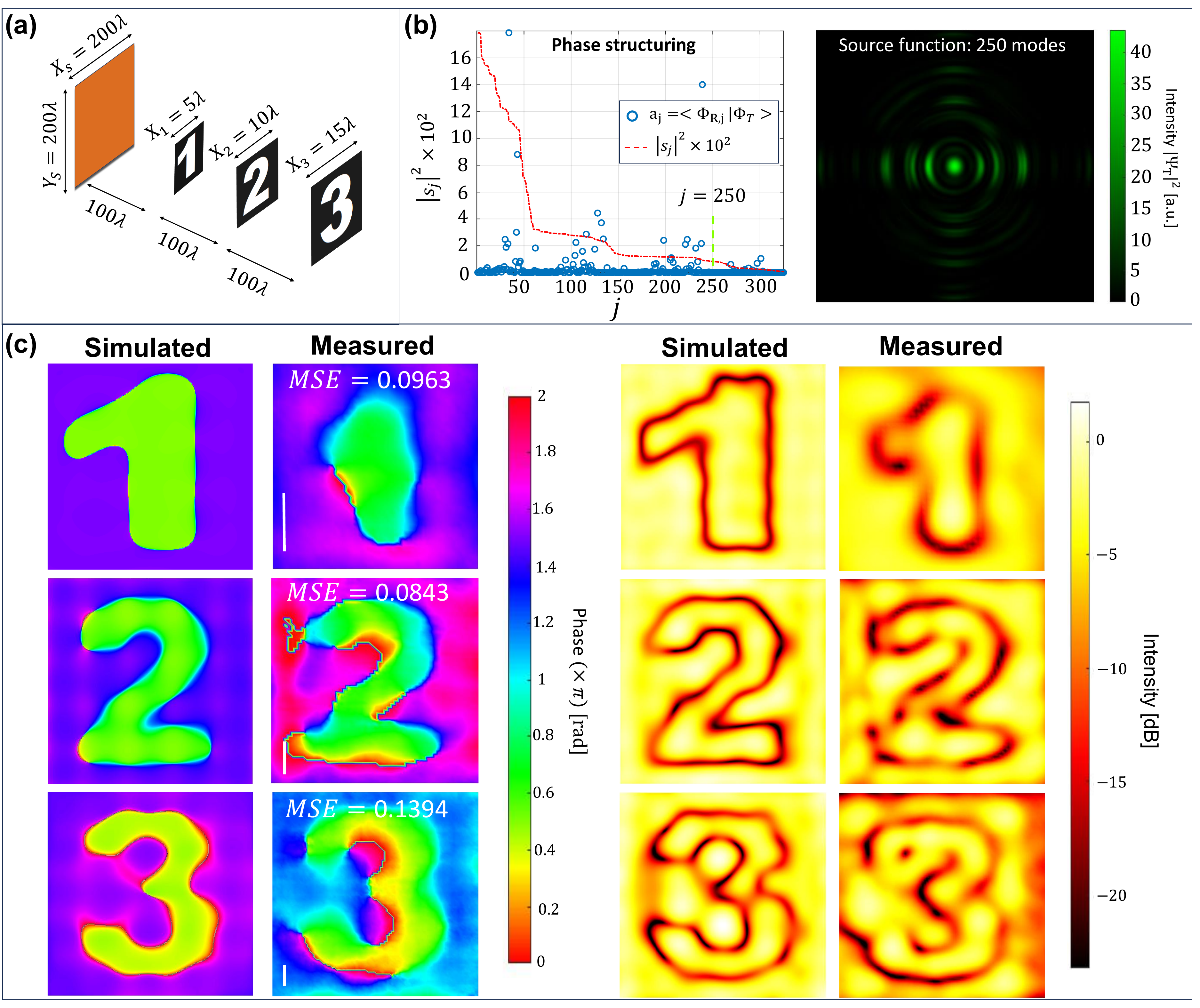}
	\caption{\small \textbf{Structuring light waves under the non-paraxial regime}. \textbf{(a)} A configuration of target planes and source aperture in this regime. \textbf{(b)} Coupling strengths (red dashed line) and projection (blue circles) of the same target phase profiles as in Fig. \ref{Fig_svd_structuring_sim}. The source function is computed from the first 250 well-coupled modes. To facilitate the measurements, the wave solution is computed at $z = 0.75L_0$ (instead of at $z = L_0$) to increase the separation distances between the measured planes. \textbf{(c)} Measured phase retrieved profiles and intensity (in dB) at the target planes. Mean squared error between measured and target phase profiles are shown in white. Scale bars (vertical white lines) represent 24 $\mu$m. The second and third planes are longitudinally spaced by 47.75 mm and 95 mm from the first plane.}
\label{Fig_non_paraxial_meas}
\end{figure}

For Fig. \ref{Fig_non_paraxial_meas} (non-paraxial case), some parameters differ. No background subtraction was executed, instead a constant value of 1 was subtracted from all pixels, to adjust the reported camera intensity range from [1,4095] to [0,4094]. Additionally, the stage positioning and $4f$ adjustment was different, as the fields encoded on the SLM were the wave solution computed at $z = 0.75L_0$, requiring a different spacing of translation stage relative to the second $4f$ lens. The collimation here was only tested visually, leading to looser alignment of the $4f$ system in the propagation direction and thus a deviation to the designed magnification.







\newpage


\begin{figure}[htbp]
	\centering	\includegraphics[width=1.0\textwidth,height=0.95\textwidth]{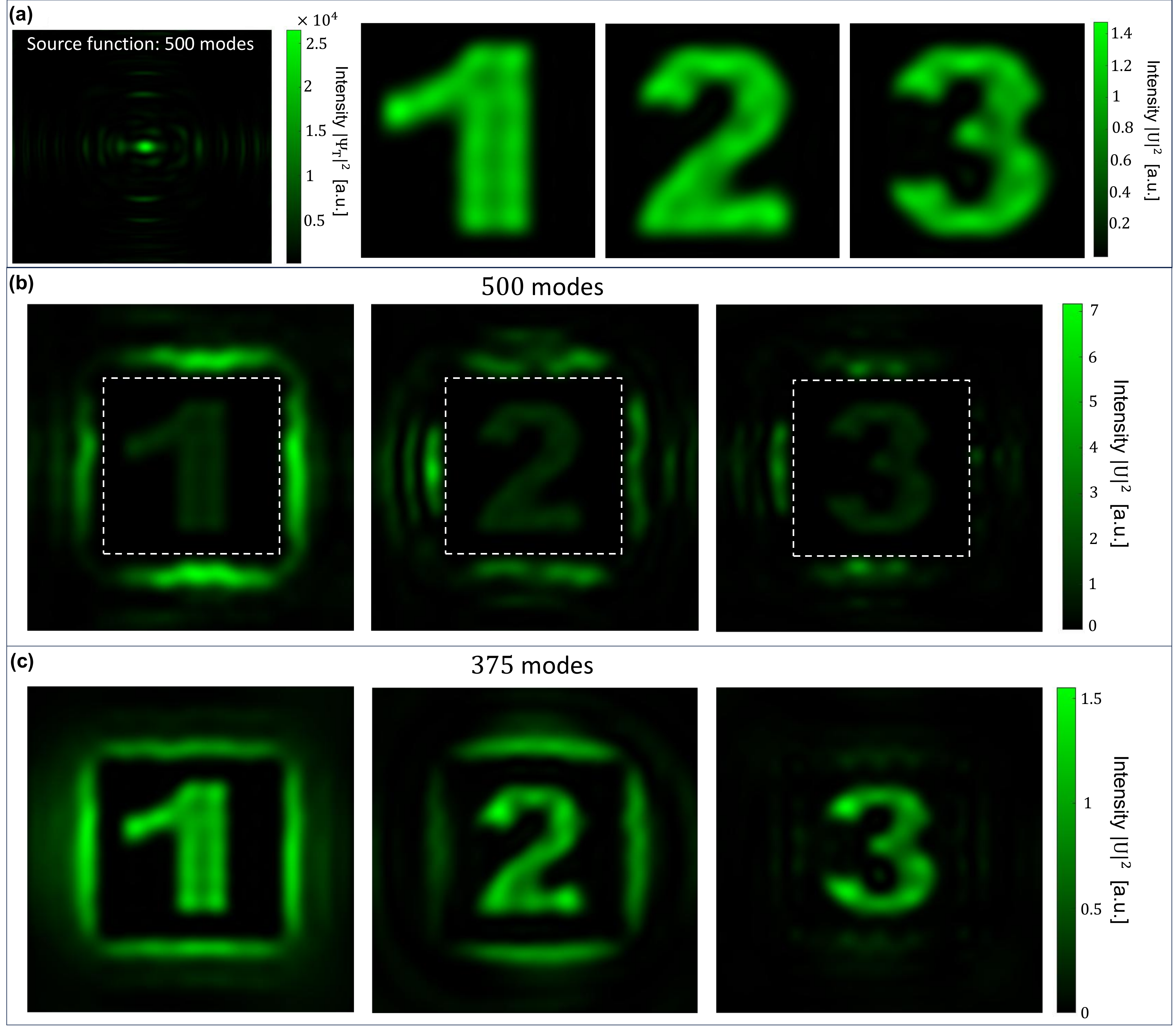}
	\caption{\small \textbf{Incorporating weakly coupled modes}. For the example of Fig. \ref{Fig_svd_structuring_sim}(b) we incorporate the first 500 modes in an attempt to fully reconstruct the flat-topped profiles of the target digits. \textbf{(a)} Calculated source function and resulting wave at the three receiving planes. The improvement in the reconstruction accuracy is afforded by a substantial increase in source amplitudes. \textbf{(b)} This high energy remains largely confined outside the target planes due to the tunneling-like escape behavior of these weakly coupled modes. The dashed white squares highlight the limits of the target planes. \textbf{(c)} For comparison, energy distribution when the first 375 modes are incorporated which includes a smaller amount of weakly coupled modes.  In this case, the intensity outside the area of interest remains about the same magnitude as the reconstructed intensity.}
\label{Fig_500_modes_int}
\end{figure}

\begin{figure}[htbp]
	\centering	\includegraphics[width=1.0\textwidth,height=0.9\textwidth]{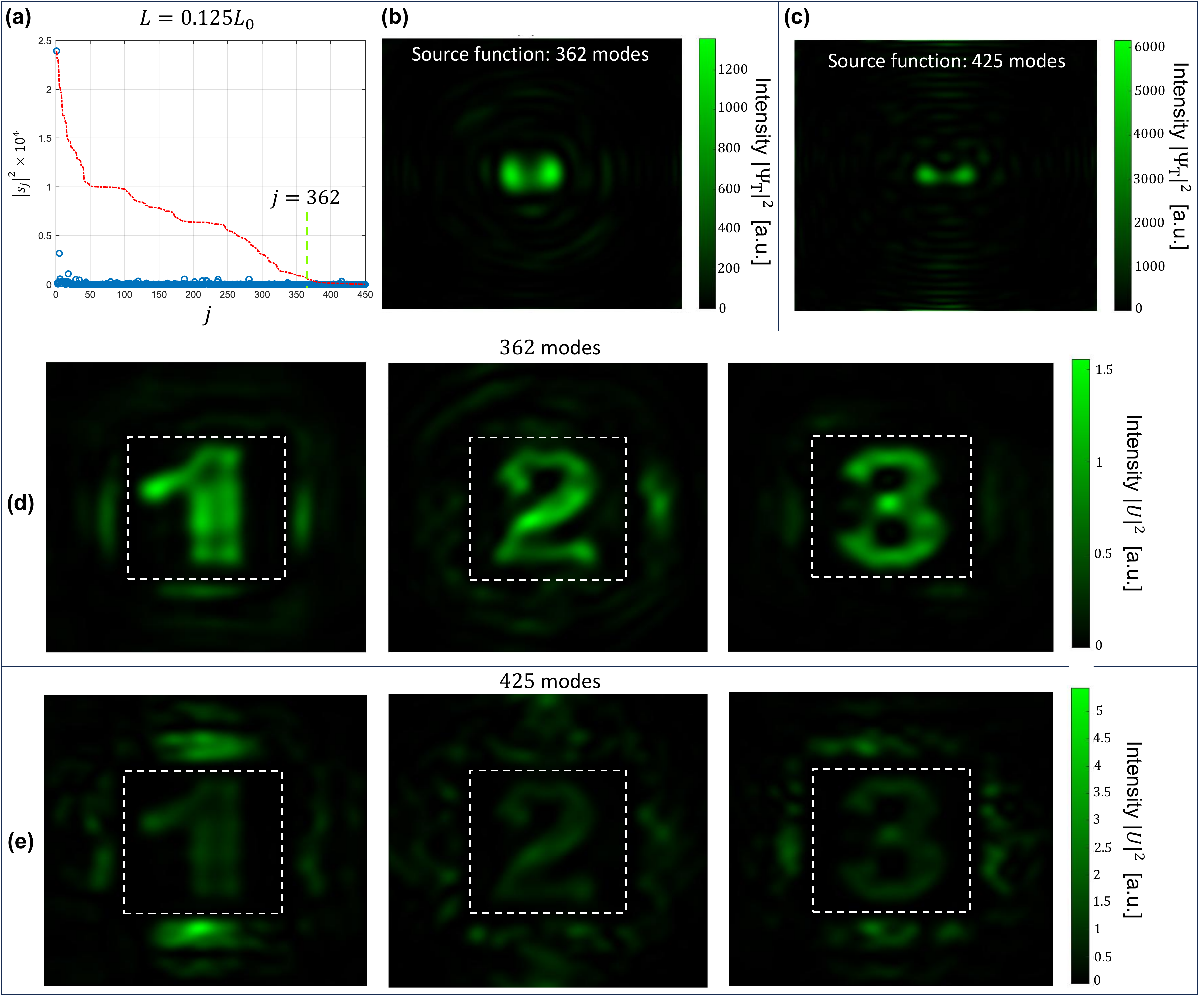}
	\caption{\small \textbf{Structuring arbitrary light wave profiles for smaller separation distances between the target planes}. \textbf{(a)} Projection of the target patterns (intensity profiles of the binary digits '1','2' and '3') onto the receiving basis (blue circles) for $L = 0.25L_0$. Calculated source function computed from the strong and partially coupled modes (first 362 modes) \textbf{(b)} and from the first 425 modes \textbf{(c)}. The set of strong and partially coupled modes is insufficient to achieve a similar reconstruction accuracy as for larger $L/L_0$ values. \textbf{(d)} Incorporating weakly coupled modes leads to a
    high-energy wave that remains mostly confined outside the target planes.}
\label{Fig_L_0_25L0}
\end{figure}

\begin{figure}[htbp]
	\centering	\includegraphics[width=0.75\textwidth,height=0.5\textwidth]{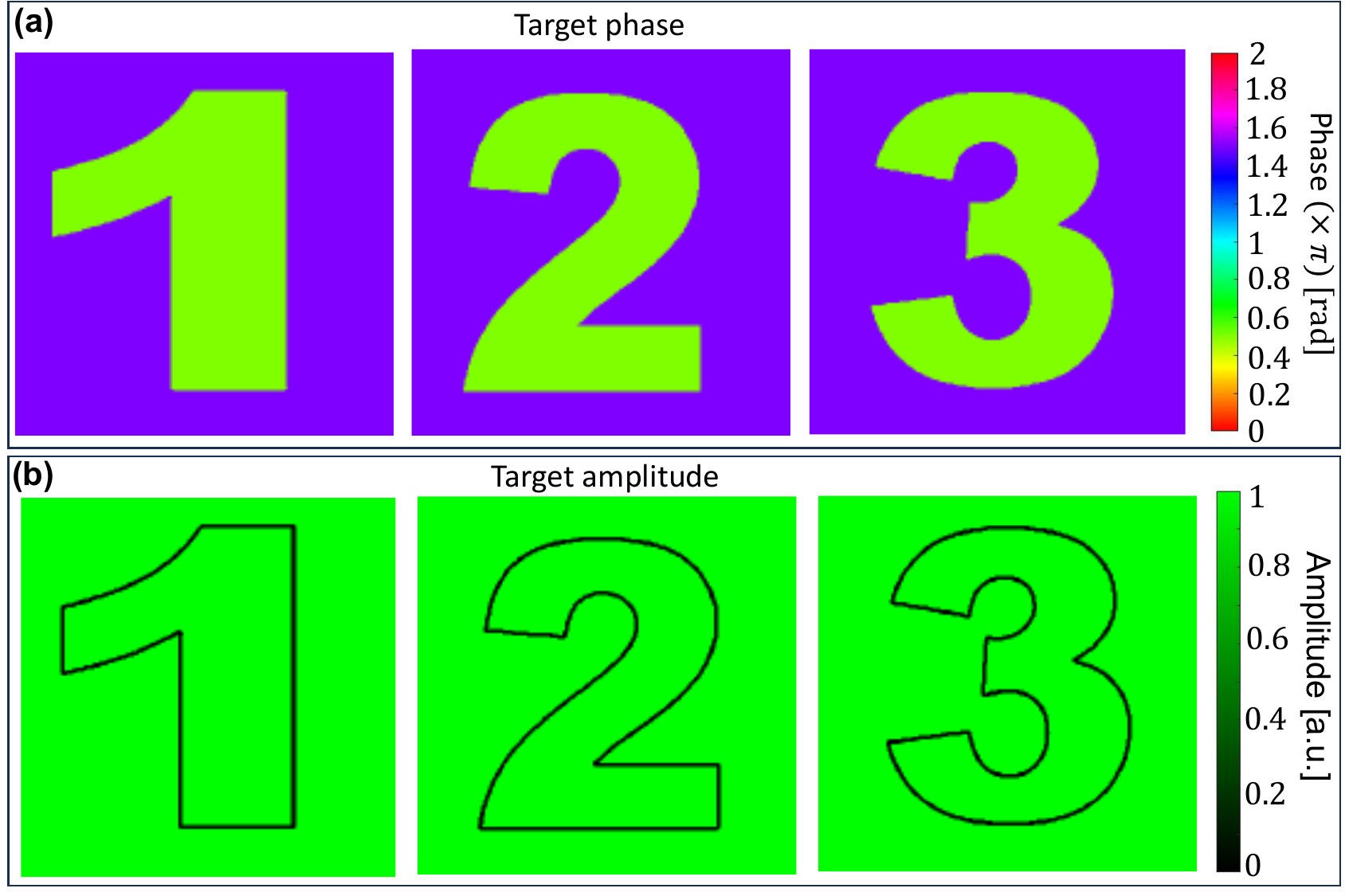}
	\caption{\small \textbf{Target phase and amplitude profiles to create 2D sheet singularities}. \textbf{(a)} A $\pi$-phase discontinuity at the prescribed singularity locations is assigned by defining a phase of $\pi/2$ radians inside the singularity contour and a phase of $3\pi/2$ radians outside.  \textbf{(b)} Along each singularity contour the amplitude is set to zero while a uniform unit value amplitude is set elsewhere.}
\label{Fig_target_phase_ampl}
\end{figure}


\begin{figure}[htbp]
	\centering	\includegraphics[width=1.0\textwidth,height=0.5\textwidth]{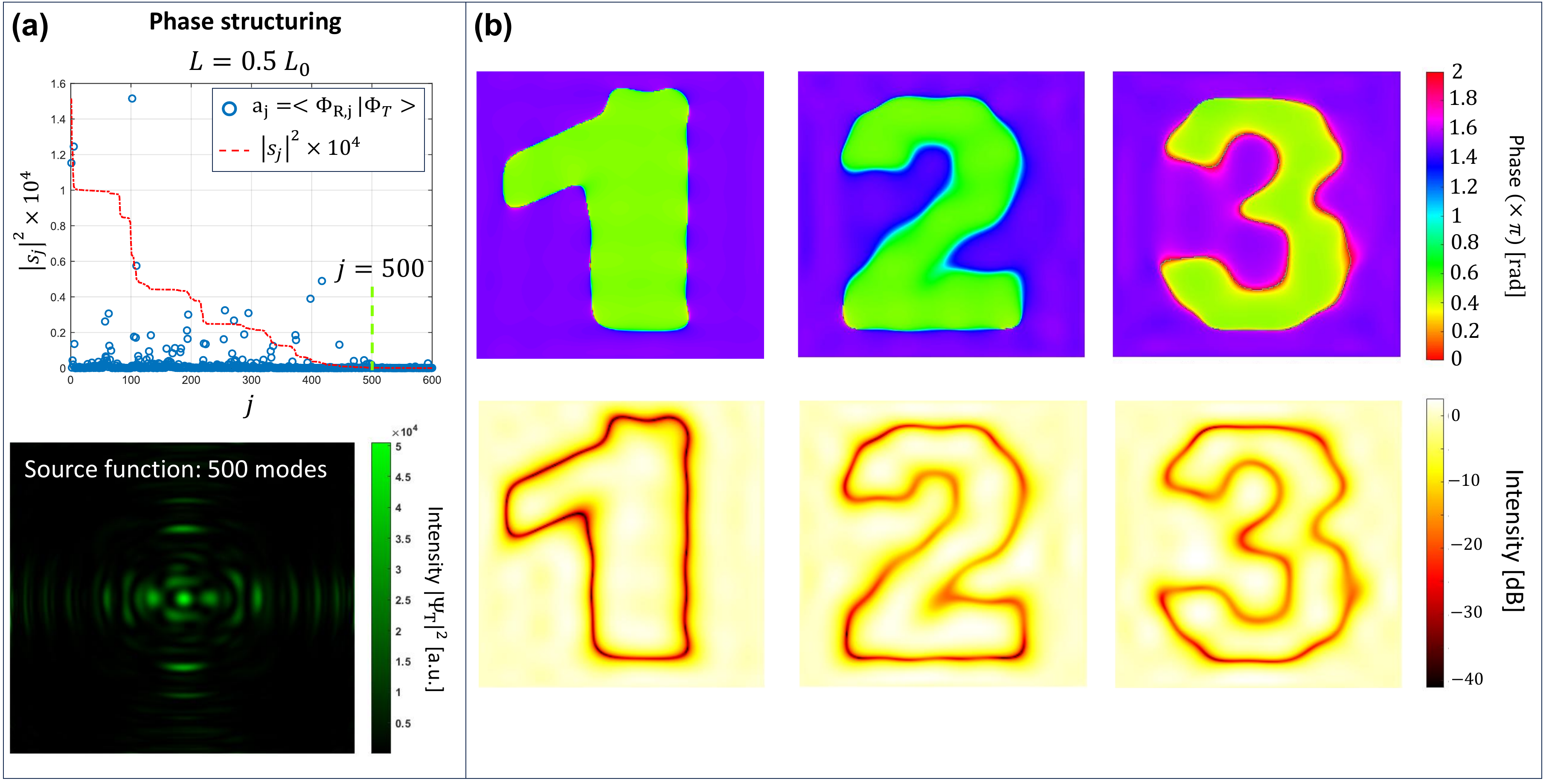}
	\caption{\small \textbf{Mathematically exact sheet phase singularities}. Due to finite source aperture, mathematically exact phase singularities can only be achieved by incorporating weakly coupled modes. \textbf{(a)} For the example of Fig. \ref{Fig_svd_structuring_sim}(e) the source function is computed from the first 500 modes. \textbf{(b)} The high-energy source function due to the incorporation of weakly coupled modes results in singular points with intensity as lower as -41 dB across all planes.}
\label{Fig_500_modes_phase}
\end{figure}

\begin{figure}[htbp]
	\centering	\includegraphics[width=1.0\textwidth,height=0.85\textwidth]{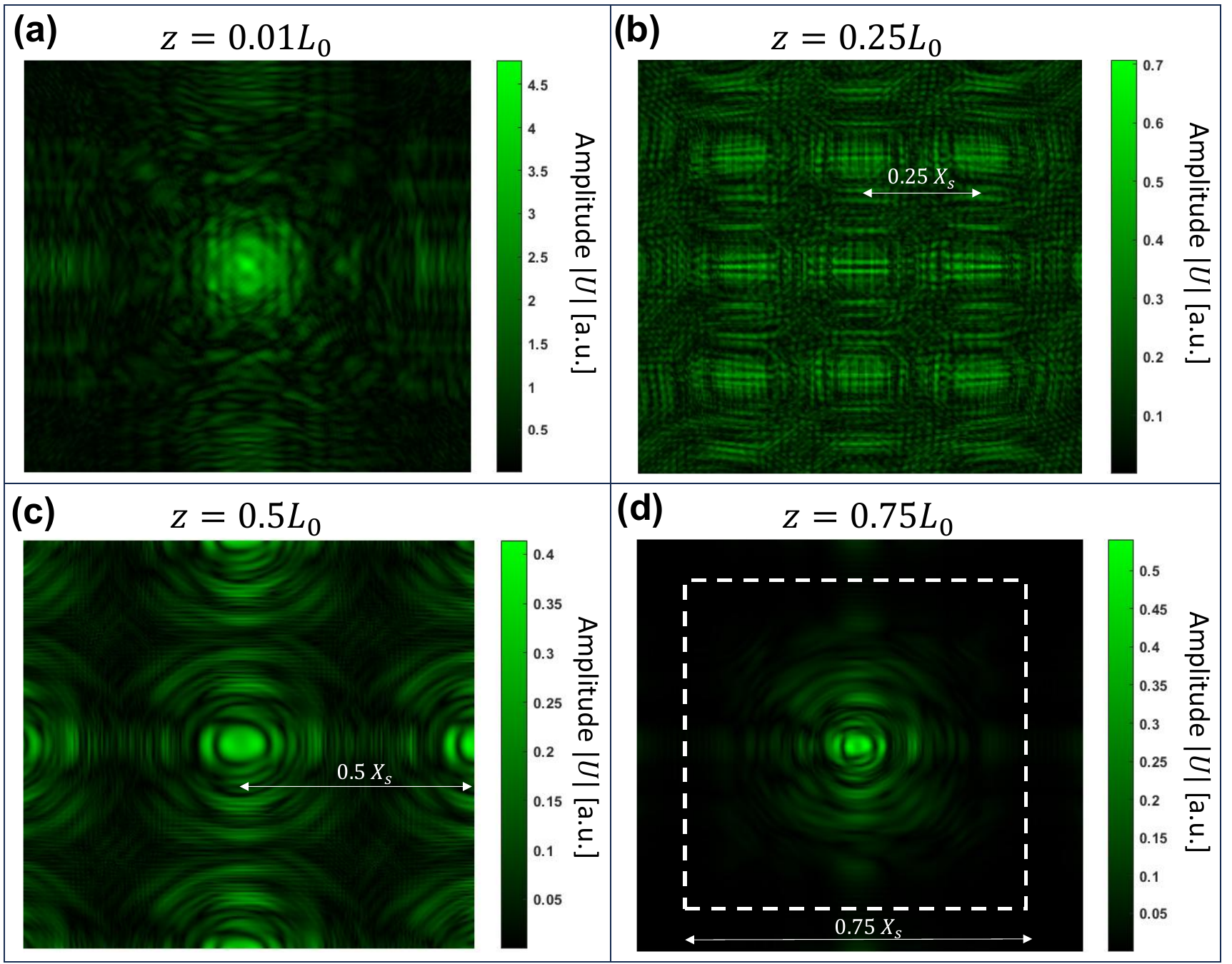}
	\caption{\small \textbf{Longitudinal evolution of the wave amplitude propagated from the source plane}. At short distances, \textbf{(a)} $z = 0.01L_0$ and \textbf{(b)} $z = 0.25L_0$, the field exhibits strong reactive near-field structure associated with evanescent spatial-frequency components. \textbf{(b)} These features decay with propagation and by \textbf{(c)} $z = 0.5L_0$ the field is mostly dominated by propagating components. However, at this distance, diffraction orders still spatially overlap with each other. \textbf{(d)} From about $z = 0.75L_0$, the wave propagated from the source plane is a smooth propagating field with no contributions from higher diffraction orders, suitable for SLM encoding. The distances highlighted in white indicates the separation distances between the diffraction orders.}
\label{Fig_reactive}
\end{figure}

\begin{figure}[htbp]
	\centering	\includegraphics[width=1.0\textwidth,height=0.35\textwidth]{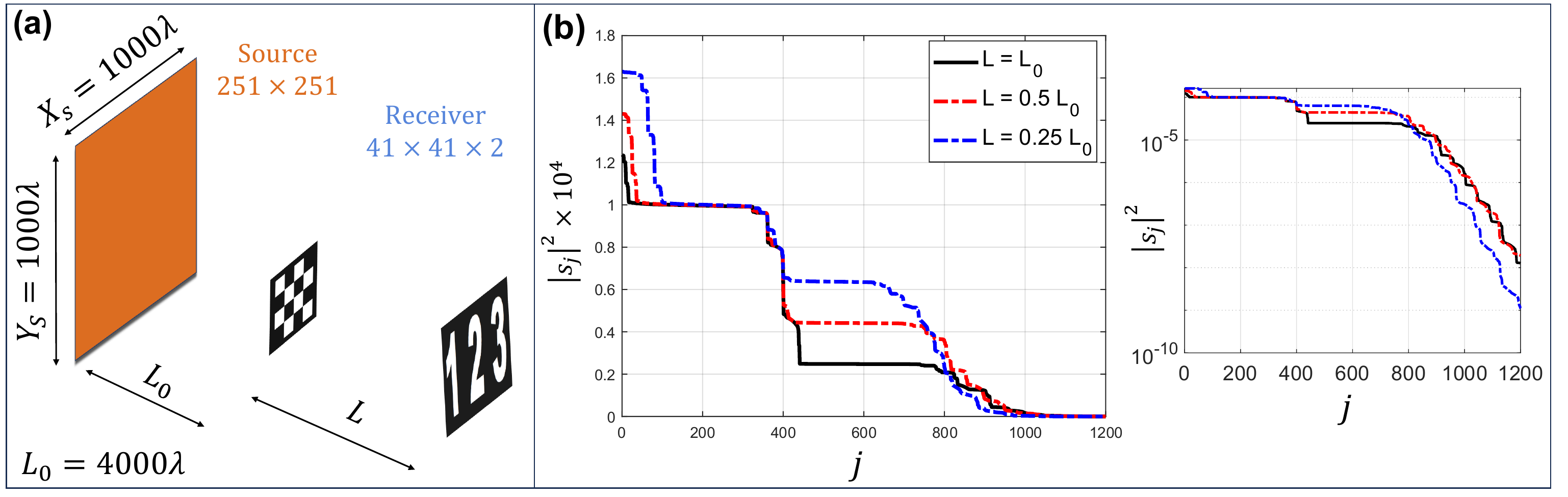}
	\caption{\small \textbf{Coupling strengths for larger receiving arrays}. \textbf{(a)} Configuration of Fig. \ref{Fig_higher_resolution}(a) analyzed for distinct values of the separation distance between the receiving planes. \textbf{(b)} Coupling strengths for distinct values of the ratio $L/L_0$ on a linear and a logarithmic scale. For these receiving arrays the total number of strongly coupled modes begins to decrease at $L = 0.25L_0$.}
\label{Fig_S_comp_L_L0_ratio}
\end{figure}


\begin{figure}[htbp]
	\centering	\includegraphics[width=1.0\textwidth,height=0.7\textwidth]{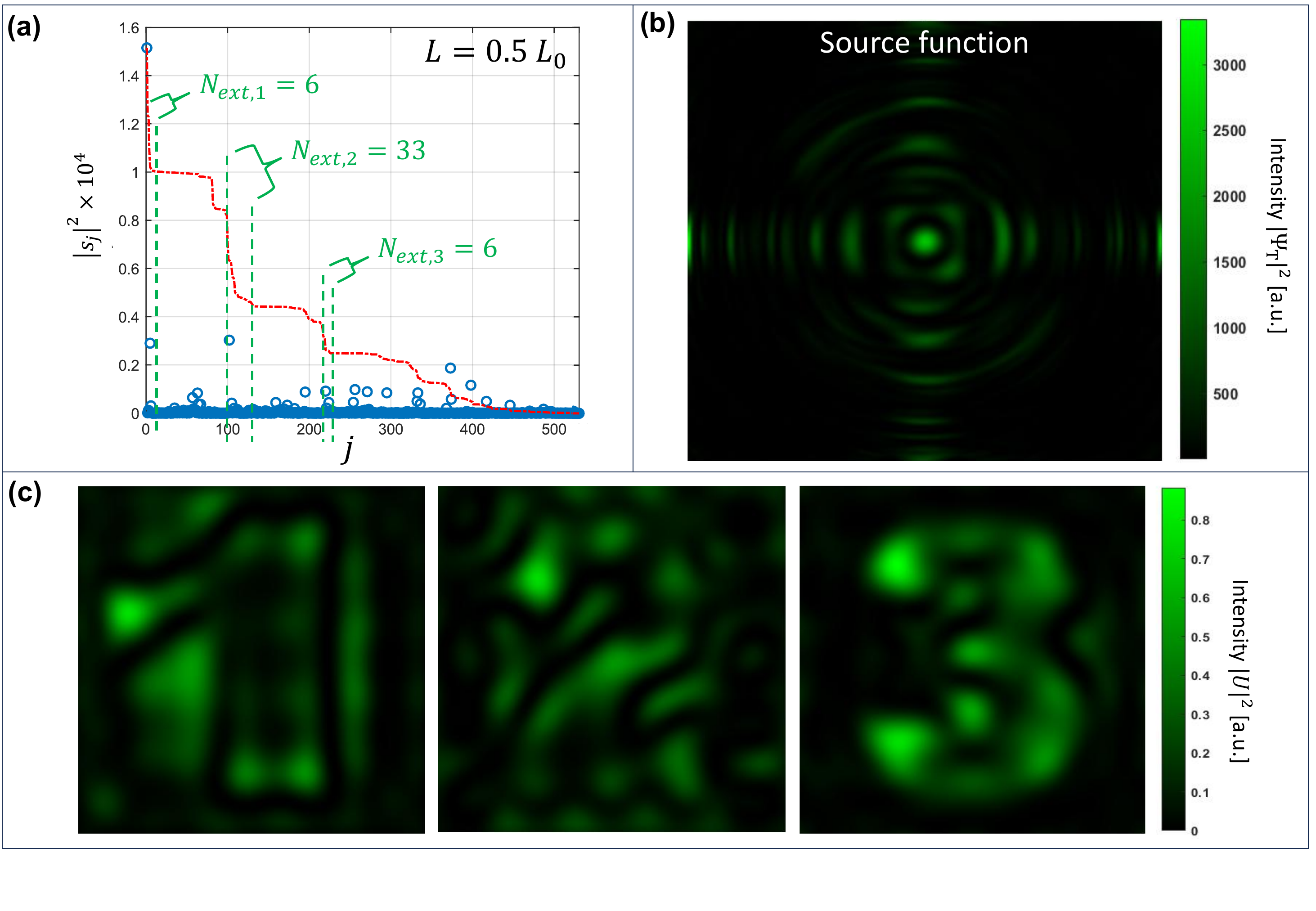}
	\caption{\small \textbf{Excluding extrinsic modes in light wave synthesis}. \textbf{(a)} For the configuration of Fig. \ref{Fig_modes_n_2}(a) with $L = 0.5 L_0$ we project the same intensity profiles as done in Fig. \ref{Fig_svd_structuring_sim}(b) - binary digits '1','2' and '3' - onto the receiving basis (blue circles). We highlight the range and number of extrinsic modes of each of the three receiving planes (green dashed lines). \textbf{(b)} When computing the source function, the extrinsic modes are not incorporated. \textbf{(c)} Resulting wave at the target planes. The degraded reconstructed profiles indicate that the extrinsic modes are crucial for suppressing inter-plane crosstalk.}
\label{Fig_remov_extr_modes}
\end{figure}

\begin{table}[ht]
\centering
\caption{Quantitative metrics for experimentally reconstructed intensity and phase profiles in Fig. \ref{Fig_higher_resolution}(d-e).}
\label{tab:metrics_higher_resolution}
\begin{tabular}{|l|lll|lll|}
\hline
      & \multicolumn{3}{l|}{Intensity structuring}                                                                            & \multicolumn{3}{l|}{Phase structuring}                                 \\ \hline
Plane & \multicolumn{1}{l|}{MSE}    & \multicolumn{1}{l|}{SBR}   & \begin{tabular}[c]{@{}l@{}}Speckle\\ Contrast\end{tabular} & \multicolumn{1}{l|}{MSE}    & \multicolumn{1}{l|}{$\sigma_{\text{int}} (\times \pi)$} & $\sigma_{\text{out}} (\times \pi)$ \\ \hline
1     & \multicolumn{1}{l|}{0.1469} & \multicolumn{1}{l|}{13.35} & 0.5338                                                     & \multicolumn{1}{l|}{0.0790} & \multicolumn{1}{l|}{0.1336}   & 0.0835   \\ \hline
2     & \multicolumn{1}{l|}{0.1086} & \multicolumn{1}{l|}{5.847} & 0.5596                                                     & \multicolumn{1}{l|}{0.2498} & \multicolumn{1}{l|}{0.2765}   & 0.5040   \\ \hline
\end{tabular}
\end{table}

\end{document}